\documentclass[nofootinbib,a4paper,12pt]{article}
\usepackage[english]{babel}
\usepackage{jheppub}
\usepackage[T1]{fontenc}
\usepackage[utf8]{inputenc}
\usepackage{float}
\usepackage{slashed}
\usepackage{multicol,multirow}
\usepackage{amssymb}
\usepackage{amsmath}
\usepackage{subcaption}
\usepackage{array}
\usepackage[dvipsnames]{xcolor}
\usepackage{hyperref}
\hypersetup{colorlinks=true}
\usepackage{longtable}
\usepackage{graphics,graphicx,epsfig,ulem,booktabs}
\numberwithin{equation}{section}
\usepackage{url}
\usepackage{physics}
\usepackage{graphicx}
\usepackage{simpler-wick}
\usepackage{feynmp}
\usepackage[compat=1.1.0]{tikz-feynman} 
\usepackage{bm}
\usepackage[fleqn]{nccmath}
\usepackage{bbold}
\usepackage{kantlipsum}
\usepackage{braket}
\usepackage{empheq}
\usepackage[most]{tcolorbox}
\usepackage{caption}
\usepackage{subcaption}
\usepackage{fancyhdr} 
\usepackage{longtable}
\usepackage{pgfplots}
\usepackage[nameinlink,capitalize]{cleveref}
\usepackage{orcidlink}

\newcommand{\GeV}{\ensuremath{\,\mathrm{GeV}}}

\definecolor{blue}{rgb}{0.2, 0.4, 1.0}
\definecolor{green}{rgb}{0.1,0.8,0.2}
\definecolor{orange}{rgb}{1.0,0.5,0.0}
\definecolor{cyan}{rgb}{0.0,0.75,0.8}

\definecolor{AM}{rgb}{0,0,1.0}
\newcommand{\alm}[1]{{\color{AM} #1}}

\newcolumntype{C}[1]{>{\centering\let\newline\\\arraybackslash\hspace{0pt}}m{#1}}

\definecolor{darkred}{RGB}{200,0,0}

\hypersetup{%
  colorlinks=true, linktocpage=true, pdfstartpage=1, pdfstartview=FitV,
  breaklinks=true, pageanchor=true,
  pdfpagemode=UseNone,
  plainpages=false, bookmarksnumbered, bookmarksopen=true, bookmarksopenlevel=1,
  hypertexnames=true, pdfhighlight=/O,
  urlcolor=RoyalBlue, linkcolor=darkred, citecolor=RoyalBlue,
  pdftitle={Enhanced Standard Model bound on $\Delta a_{\rm CP}^{\rm dir}$ from QCD penguins},
  pdfauthor={Ali Mohamed, Maria Laura Piscopo},
  pdfsubject={},
  pdfkeywords={},
  pdfcreator={pdfLaTeX},
  pdfproducer={LaTeX with hyperref}
}

\title{\boldmath 
Enhanced Standard Model bound on charm CP violation from QCD penguins}

\preprint{
\small
\begin{tabular}{l}
P3H-26-072\\
SI-HEP-2026-22\\
CERN-TH-2026-220
\end{tabular}
}

\author[a]{Ali Mohamed\orcidlink{0009-0006-3466-2709}}
\author[b]{, Maria Laura Piscopo\orcidlink{0000-0003-0796-5561}}

\affiliation[a]{Physik Department, Universit\"{a}t Siegen, Walter-Flex-Str. 3, D-57068 Siegen, Germany}
\affiliation[b]{Theoretical Physics Department, CERN, 1211 Geneva 23, Switzerland}

\emailAdd{ali.mohamed@uni-siegen.de}
\emailAdd{maria.laura.piscopo@cern.ch}

\abstract{
We investigate the impact of QCD penguin operators on direct CP violation in singly Cabibbo-suppressed $D^0$ decays within the framework of light-cone sum rules (LCSR) using pion and kaon light-cone distribution amplitudes. 
We determine, for the first time, the hadronic matrix elements of the QCD penguin operators for $D^0\to K^+K^-$ and $D^0\to\pi^+\pi^-$ at leading order in $\alpha_s$, employing a three-point correlation function with an artificial momentum to avoid unphysical parasitic cuts. We also compute the matrix elements of the current-current operators using the same framework and compare them with previous LCSR results. We find that the scalar penguin operators $Q_5$ and $Q_6$ can have substantially enhanced matrix elements.
We identify two sources of enhancement: quark-condensate contributions, which arise already at tree level and are therefore enhanced by a relative factor of $16 \pi^2$ compared to the remaining, loop-induced terms in the sum rule; and an annihilation-type contribution from the $\bar u u$ component of these operators, which enters already through twist-three light-cone distribution amplitudes and is of comparable magnitude.
While the impact of QCD penguin operators on the branching ratios is negligible, they can enhance the magnitude of the ``penguin-to-tree'' amplitude ratios relevant for direct CP violation. The relevant strong phases, however, remain largely unconstrained within the accuracy of our framework. 
Assuming maximal relative strong phases, we find that the corresponding upper bound on the Standard Model contribution to charm CP violation can be enhanced compared to estimates based on the current-current operators alone, reducing the gap with the experimental measurement.
}

\begin{document}

\maketitle

\section{Introduction}
The observation by the LHCb collaboration in 2019 of a non-zero difference between the CP asymmetries in the singly Cabibbo-suppressed decays
$D^0 \to K^+K^-$ and $D^0 \to\pi^+\pi^-$, denoted by $\Delta A_{\rm CP}$,
opened a puzzle in flavour physics that remains unresolved.
The measured asymmetry, at the level of $10^{-3}$~\cite{LHCb:2019hro}, is intriguing because CP violation in the charm sector is suppressed in the Standard Model (SM) by the small weak factor
${\rm Im}(\lambda_b/\lambda_d)\sim 10^{-3}$~\cite{ParticleDataGroup:2026aaa}, where $\lambda_q\equiv V_{cq}^*V_{uq}$ denote the relevant combinations of
Cabibbo-Kobayashi-Maskawa (CKM) matrix elements. 
The size of the observed effect therefore points to an enhancement of the relevant ``penguin-to-tree'' amplitude ratios to values of order unity, roughly an order of magnitude larger than suggested by naive perturbative estimates, see e.g.~\cite{Grossman:2006jg}. 
The theoretical uncertainties associated with these hadronic quantities, however, are large enough that the measurement cannot be unambiguously interpreted as evidence for physics beyond the SM~(BSM).
Although possible BSM explanations have been investigated~\cite{Chala:2019fdb, Dery:2019ysp, Calibbi:2019bay, Bause:2020obd}, a quantitative assessment of whether the observed size of CP violation can be accommodated within the SM ultimately requires a solid understanding of the underlying non-perturbative QCD dynamics.
This is particularly challenging in charm decays, see e.g.~\cite{Friday:2025gpj, Lenz:2020awd} for recent reviews. The relatively close value of the charm-quark mass, $m_c\sim 1$~GeV, to the typical hadronic scale $\Lambda_{\rm QCD}$ of a few hundred MeV implies potentially sizeable perturbative and power corrections, while robust predictions for non-leptonic decays are notoriously difficult to obtain even in the more favourable case of $B$-meson decays.

Various approaches have been employed to address this problem.
An enhancement of the SM contribution to $\Delta A_{\rm CP}$ was proposed in~\cite{Grossman:2019xcj}, based on the $U$-spin decomposition of the decay amplitudes and a mechanism analogous to the $\Delta I=1/2$ rule in kaon decays. 
However, questions about the robustness of invoking $U$-spin symmetry in charm decays have been raised 
following more precise experimental determinations of the CP asymmetries in
$D^0\to K^+K^-$ and $D^0\to\pi^+\pi^-$ by the LHCb collaboration~\cite{LHCb:2022lry}, which indicate a departure of about 2.7$\sigma$ from the strict $U$-spin-symmetric limit.~\footnote{This tension is reduced when $U$-spin breaking in the corresponding decay amplitudes is taken into account, see e.g.,~\cite{LHCb:2024yxi, PajeroatCharm}.}
The implications of these experimental results have been discussed in~\cite{Schacht:2022kuj, Bause:2022jes, Iguro:2024uuw, Gavrilova:2023fzy, Sinha:2025cuo}, while suitably constructed sum rules have been proposed to further test the size of $U$-spin breaking in the charm system~\cite{Gavrilova:2026ryc, Iguro:2024uuw}.

Global fits based on topological-amplitude decompositions are also widely used to gain information on the possible size of the relevant hadronic contributions~\cite{Cheng:2019ggx, Bolognani:2025rvk, Li:2019hho, Wang:2020gmn}.
More recently, it was shown in~\cite{Fleischer:2025zhl} that combining experimental information on branching ratios across the singly Cabibbo-suppressed channels
$D^0\to K^+K^-$, $D^0\to\pi^+\pi^-$ and $D^0\to K_{\rm S}^0K_{\rm S}^0$ with precise lattice QCD input provides an interesting avenue for testing factorisation and deriving benchmark scenarios for the CP asymmetry in $D^0\to K_{\rm S}^0K_{\rm S}^0$ to be tested against future experimental data. 

More quantitative studies of the non-perturbative dynamics underlying $D^0 \to \pi^+\pi^-$ and $D^0 \to K^+K^-$ have been carried out using dispersive methods combined with a data-driven approach~\cite{Pich:2023kim, Pich:2026zux, ValeSilva:2024vmv}. These analyses exploit experimental information on $\pi\pi \to K K$ rescattering to estimate final-state interactions and, importantly, to constrain the relevant strong phases. So far, it was found that rescattering effects alone appear insufficient to account for the experimental value of $\Delta A_{\rm CP}$. This conclusion differs from that of~\cite{Bediaga:2022sxw}, where a study of final-state interactions found that rescattering could substantially enhance the SM prediction. Possible origins of this discrepancy have been discussed in~\cite{Pich:2023kim}.
On the other hand, large rescattering contributions, particularly those associated with scalar resonances close to the $D$ meson mass, such as $f_0(1710)$ and $f_0(1790)$, have been suggested in~\cite{Schacht:2021jaz, Soni:2019xko}, pointing at a possible large effect from scalar operators entering the weak Hamiltonian.

The framework of light-cone sum rules (LCSR)~\cite{Balitsky:1989ry} provides a complementary approach to determining the hadronic matrix elements relevant for charm CP violation. Building on the methodology developed for $B\to\pi\pi$ in~\cite{Khodjamirian:2003eq}, LCSR was first applied in~\cite{Khodjamirian:2017zdu} to compute the penguin contractions of the current-current operators entering the effective Hamiltonian for $D^0\to\pi^+\pi^-$ and $D^0\to K^+K^-$. These contributions arise at ${\cal O}(\alpha_s)$ and generate non-trivial strong phases. More recently, LCSR was employed to determine the factorisable tree-level matrix elements of the current-current operators~\cite{Lenz:2023rlq}, which arise at ${\cal O}(\alpha_s^0)$ and therefore do not generate an additional strong phase at this order. In both analyses, the resulting contributions to direct CP violation were found to be insufficient to account for the experimentally observed value.

So far, these LCSR analyses have focused exclusively on current-current operators, while the contributions of QCD penguin operators have been neglected because of their small Wilson coefficients.
However, the penguin contractions of the current-current operators and the contributions of the QCD penguin operators to the decay amplitudes are both formally ${\cal O}(\alpha_s)$. 
Their inclusion should therefore be taken into account for a consistent treatment of the renormalisation-scale and scheme dependence associated with the matching of Wilson coefficients and hadronic matrix elements, see e.g.~\cite{Buras:1997cv, Buras:1998ra}.
Moreover, the scalar QCD penguin operators $Q_5$ and $Q_6$ are chirally enhanced and receive potentially sizeable annihilation contributions~\cite{Gronau:1999zt}. These effects may compensate, at least in part, for the suppression associated with their Wilson coefficients. The possibility of enhanced penguin contributions at the charm scale was also discussed in~\cite{Brod:2011re,Brod:2012ud}. Nevertheless, a non-perturbative determination of the matrix elements of the QCD penguin operators is still lacking.

In this work, we take a first step towards filling this gap by determining, for the first time, the matrix elements of the QCD penguin operators for $D^0 \to K^+K^-$ and $D^0 \to\pi^+\pi^-$ within LCSR at leading order in $\alpha_s$. Specifically, we follow the approach developed in the study of non-leptonic $B\to\pi\pi$ decays in~\cite{Khodjamirian:2000mi} and subsequently applied to annihilation contributions in these modes in~\cite{Khodjamirian:2005wn}, adapting it here to the charm sector. The method is based on a three-point correlation function with on-shell pion and kaon states and employs a light-cone operator product expansion~(LC-OPE) together with hadronic dispersion relations to extract the desired matrix elements. In particular, following~\cite{Khodjamirian:2000mi}, an artificial momentum is introduced into the correlation function to avoid contamination of the dispersion relations by unphysical parasitic cuts.
This approach therefore differs from that previously adopted in~\cite{Lenz:2023rlq}, where no artificial momentum was introduced. We investigate the differences between the two approaches by computing the matrix elements of the current-current operators and comparing our results with those of the previous determination.

We find that the matrix elements of the scalar penguin operators $Q_5$ and $Q_6$ can be substantially larger than those of the current-current operators. Within our framework, this enhancement has two sources: sizeable quark-condensate contributions, which arise already at tree level and are therefore enhanced by a relative factor of $16 \pi^2$ compared to the remaining, loop-induced terms in the sum rule, and an annihilation-type contribution associated with the $\bar u u$ component of the scalar operators. Since our analysis is performed at leading order in $\alpha_s$, however, it does not provide additional information on the relative strong phases.

Despite the enhancement of their hadronic matrix elements, the QCD penguin contributions remain negligible for the branching ratios because of their small Wilson coefficients and CKM suppression. They can nevertheless have a significant impact on the ``penguin-to-tree'' amplitude ratios and, consequently, on the possible size of direct CP violation. Assuming maximal relative strong phases, we find that the resulting upper bound on the SM contribution to direct CP violation, when uncertainties are taken into account, can be larger than in previous LCSR studies, reducing the gap with the experimental measurement.
Within the adopted framework, however, the limited control over the relevant strong phases and higher-order perturbative and power corrections prevents us from obtaining quantitative predictions. Nevertheless, our results indicate that QCD penguin matrix elements may play an important role in assessing the size of the SM contribution and motivate further dedicated studies of these effects.

The rest of the paper is organised as follows. We start by introducing the weak effective framework for the study of the singly Cabibbo-suppressed decays $D^0 \to \pi^+ \pi^-$ and $D^0 \to K^+ K^-$ as well as the relevant observables in section~\ref{sec:Heff}. The description of the LCSR method used to determine the hadronic matrix elements of the current-current and penguin operators is presented in section~\ref{sec:LCSR}, while details of the calculation of the corresponding LC-OPE are discussed in section~\ref{sec:OPE}. 
Section~\ref{sec:Results} contains the numerical analysis and our results for the hadronic matrix elements of the current-current and QCD penguin operators together with their phenomenological implications on the expected size of direct CP violation. Finally, we summarise our results and conclude in section~\ref{sec:conclusion}. 
%%%%%%%%%%%%%%%%%%%%%%%%%%%%%%%%%%%%%%%%%%%%%%%%
%%%%%%%%%%%%%%%%%%%%%%%%%%%%%%%%%%%%%%%%%%%%%%%%
%%%%%%%%%%%%%%%%%%%%%%%%%%%%%%%%%%%%%%%%%%%%%%%%
\section{Decay amplitudes within the weak effective theory}
\label{sec:Heff}
The starting point for the analysis of the singly Cabibbo-suppressed  decays $D^0\to\pi^+ \pi^-$
and $D^0\to K^+K^-$ is the $\Delta C=1$ weak effective
Hamiltonian describing the underlying flavour-changing charm-quark transitions
$c\to q\bar qu$, with $q=u,d,s$~\cite{Buchalla:1995vs}:
\begin{equation}
{\cal H}_{\rm eff} = \frac{G_F}{\sqrt{2}} \Bigg[ \sum_{q = d,s} \lambda_{q}\left(C_1 Q_1^{q} + C_2 Q_2^{q}\right) - \lambda_b \sum_{k = 3}^6 C_k Q_k \Bigg] + {\rm h.c.} \,
\label{eq:Heff}
\end{equation} 
Here, $G_F$ is the Fermi constant, $\lambda_q\equiv V_{cq}^*V_{uq}$ denote the relevant combinations of
CKM matrix elements, $Q_1^q$ and $Q_2^q$ are the current-current operators,
\begin{align}
Q_1^q = \left(\bar q^i   c^i \right)_{V-A}  \left(\bar u^j q^j \right)_{V-A}\,, \qquad 
Q_2^q = \left(\bar q^i  c^j \right)_{V-A}  \left(\bar u^j  q^i \right)_{V-A}\,,
\label{eq:Q1}
\end{align} 
whereas the QCD penguin operators are
\begin{equation}
Q_k \equiv \sum_{q = u,d,s} Q_k^q\,, \qquad k = 3, \ldots, 6\,,
\label{eq:QC_P}
\end{equation}
with
\begin{align}
Q^q_3 = \left(\bar u^i c^i \right)_{V-A}  \left(\bar q^j  q^j \right)_{V-A}\,, 
\qquad 
Q^q_4 = \left(\bar u^i   c^j \right)_{V-A}  \left(\bar q^j  q^i \right)_{V-A}\,,
\label{eq:Q3}
\\[2mm]
Q^q_5 = \left(\bar u^i  c^i \right)_{V-A} \left(\bar q^j   q^j \right)_{V+A}\,, 
\qquad
Q^q_6 = \left(\bar u^i  c^j \right)_{V-A}  \left(\bar q^j   q^i \right)_{V+A}\,.
\label{eq:Q5}
\end{align}
In the above equations, $i,j$ indicate colour indices and we have introduced the shorthand notation $(\bar q_1 q_2)_{V\pm A}\equiv \bar q_1\gamma_\mu(1\pm\gamma_5)q_2$ where contraction over the Lorentz index is understood.
In eq.~\eqref{eq:Heff}, $C_i(\mu)$, with $i=1,\ldots,6$, are the Wilson coefficients evaluated at the renormalisation scale $\mu \sim m_c$. 
In table~\ref{tab:wc}, we collect their numerical values for different choices of $\mu$, at leading order (LO) and next-to-leading order (NLO) in QCD~\cite{Buchalla:1995vs}.

\begin{table}[t]
    \centering
    \renewcommand{\arraystretch}{1.5}
    \begin{tabular}{c|c|c|c|c|c|c}
    \toprule
    $\mu~[\mathrm{GeV}]$
    & $C_1(\mu)$
    & $C_2(\mu)$
    & $C_3(\mu)$
    & $C_4(\mu)$
    & $C_5(\mu)$
    & $C_6(\mu)$ \\
    \hline

    1 &
    $\begin{array}{c}
        1.262\\[-3mm]
        (1.352)
    \end{array}$ &
    $\begin{array}{c}
        -0.494\\[-3mm]
        (-0.645)
    \end{array}$ &
    $\begin{array}{c}
        0.025\\[-3mm]
        (0.017)
    \end{array}$ &
    $\begin{array}{c}
        -0.061\\[-3mm]
        (-0.037)
    \end{array}$ &
    $\begin{array}{c}
        0.010\\[-3mm]
        (0.011)
    \end{array}$ &
    $\begin{array}{c}
        -0.072\\[-3mm]
        (-0.047)
    \end{array}$ \\
    \hline

    1.5 &
    $\begin{array}{c}
        1.181\\[-3mm]
        (1.243)
    \end{array}$ &
    $\begin{array}{c}
        -0.369\\[-3mm]
        (-0.482)
    \end{array}$ &
    $\begin{array}{c}
        0.014\\[-3mm]
        (0.009)
    \end{array}$ &
    $\begin{array}{c}
        -0.037\\[-3mm]
        (-0.021)
    \end{array}$ &
    $\begin{array}{c}
        0.008\\[-3mm]
        (0.007)
    \end{array}$ &
    $\begin{array}{c}
        -0.039\\[-3mm]
        (-0.025)
    \end{array}$ \\
    \hline

    2 &
    $\begin{array}{c}
        1.144\\[-3mm]
        (1.195)
    \end{array}$ &
    $\begin{array}{c}
        -0.306\\[-3mm]
        (-0.404)
    \end{array}$ &
    $\begin{array}{c}
        0.009\\[-3mm]
        (0.005)
    \end{array}$ &
    $\begin{array}{c}
        -0.025\\[-3mm]
        (-0.014)
    \end{array}$ &
    $\begin{array}{c}
        0.006\\[-3mm]
        (0.004)
    \end{array}$ &
    $\begin{array}{c}
        -0.026\\[-3mm]
        (-0.015)
    \end{array}$ \\
    \bottomrule
    \end{tabular}
    \caption{Comparison of NLO (LO) values of the $\Delta C = 1$ Wilson coefficients for different choices of the renormalisation scale $\mu$~\cite{Buchalla:1995vs}.
    }
    \label{tab:wc}
\end{table}

To make the CKM structure of the decay amplitudes more transparent, following~\cite{Khodjamirian:2017zdu, Lenz:2023rlq}, we introduce a compact notation for the combination of the effective operators and their Wilson coefficients entering eq.~\eqref{eq:Heff}. Specifically, for the current-current operators, we define:
\begin{equation}
    {\cal O}^{q} \equiv - \frac{G_F}{\sqrt 2}\sum_{k=1,2}C_k Q_k^{q}\,, \qquad \mbox{with } q = d,s\,,
\label{eq:Oq-def}
\end{equation}
and for the QCD penguin operators:
\begin{equation}
    {\cal P} \equiv - \frac{G_F}{\sqrt 2}\sum_{k=3}^6 C_k  Q_k\,.
\label{eq:Op-def}
\end{equation}
With these definitions, the decay amplitudes take the compact form
\begin{align}
    {\cal A}(D^0 \to \pi^+ \pi^-) &= \lambda_d \langle \pi^+ \pi^-| {\cal O}^d| D^0 \rangle + \lambda_s \langle \pi^+ \pi^-| {\cal O}^s| D^0 \rangle - \lambda_b \langle \pi^+ \pi^-| {\cal P}| D^0 \rangle  \,,
    \label{eq:Am-pipi}
    \\[2mm]
   {\cal A}(D^0 \to K^+ K^-) &= \lambda_s \langle K^+ K^-| {\cal O}^s| D^0 \rangle + \lambda_d \langle K^+ K^-| {\cal O}^d| D^0 \rangle - \lambda_b \langle K^+ K^-| {\cal P}| D^0 \rangle\,.
    \label{eq:Am-KK}
\end{align}
\noindent
Next, using the unitarity of the CKM matrix  $\lambda_d + \lambda_s + \lambda_b = 0$, to eliminate $\lambda_s$ from eq.~\eqref{eq:Am-pipi} and $\lambda_d$ from eq.~\eqref{eq:Am-KK}, we obtain
\begin{align}
    {\cal A}(D^0 \to \pi^+ \pi^-) & =  \lambda_d \, {\cal A}_{\pi \pi} \left[1 - \frac{\lambda_b}{\lambda_d} \frac{{\cal P}_{\pi\pi}}{{\cal A }_{\pi \pi}} \right] \,,
    \label{eq:A-Dpipi}
    \\[2mm]
   {\cal A}(D^0 \to K^+ K^-) &= \lambda_s \, {\cal A}_{K K} \left[1 - \frac{\lambda_b}{\lambda_s} \frac{{\cal P}_{KK}}{{\cal A }_{KK}} \right]\,,
    \label{eq:A-DKK}
\end{align}
where we have defined, respectively,
\begin{align}
{\cal A}_{\pi\pi} & \equiv \langle \pi^+ \pi^-| {\cal O}^d| D^0 \rangle - \langle \pi^+ \pi^-| {\cal O}^s| D^0 \rangle\,,
 \label{eq:Apipi}
\\[2mm]
{\cal A}_{KK} & \equiv \langle K^+ K^-| {\cal O}^s| D^0 \rangle - \langle K^+ K^-| {\cal O}^d| D^0 \rangle\,,
 \label{eq:Akk}
\end{align}
and 
\begin{equation}
    {\cal P}_{\pi\pi} \equiv \langle \pi^+ \pi^-| {\cal O}^s + {\cal P}| D^0 \rangle\,, \qquad   {\cal P}_{KK} \equiv \langle K^+ K^-| {\cal O}^d + {\cal P} | D^0 \rangle\,.
    \label{eq:P}
\end{equation}
The expressions in eqs.~\eqref{eq:A-Dpipi}, \eqref{eq:A-DKK} make explicit the CKM hierarchy between the terms proportional to the leading CKM factors $\lambda_{d}, \lambda_s$ and the subleading factor $\lambda_b$, owing to
$|\lambda_b/\lambda_{d}| \approx |\lambda_b/\lambda_{s}|= {\cal O}(10^{-3})$~\cite{Charles:2004jd}.  
Up to corrections proportional to $\mbox{Re}(\lambda_{b}/\lambda_{s,d}) = {\cal O}( 10^{-4})$, the CP-averaged branching ratios therefore take the simple form
\begin{equation}
{\cal B} (D^0 \to \pi^+ \pi^-) =  {\cal N}_{\pi\pi} |\lambda_d|^2 |{\cal A}_{\pi\pi}|^2 \,,
\qquad
{\cal B} (D^0 \to K^+ K^-) = {\cal N}_{KK}  |\lambda_s|^2 |{\cal A}_{KK}|^2 \,,
\label{eq:Br}
\end{equation}
where ${\cal N}_{\pi\pi}$ and ${\cal N}_{KK}$ denote the corresponding phase-space normalisation factors, e.g.
\begin{equation}
    {\cal N}_{\pi\pi} = \frac{\sqrt{\lambda(m_D^2, m_\pi^2, m_\pi^2)}}{16 \pi m_D^3} \tau(D^0)\,,
\end{equation}
with $\lambda(a,b,c) \equiv (a-b-c)^2 - 4 bc$ being the Källén function and $\tau(D^0)$ the $D$-meson lifetime.

Direct CP violation in $D^0 \to f$, with $f =\{ \pi^+ \pi^-, K^+ K^-\}$, is defined as
\begin{equation}
a_{\rm CP}^{\rm dir} (D^0 \to f) \equiv \frac{|{\cal A}(D^0 \to f)|^2 - |{\cal A}(\bar D^0 \to f)|^2}{|{\cal A}(D^0 \to f)|^2 + |{\cal A}(\bar D^0 \to f)|^2}\,,
\end{equation}
and a non-zero value necessarily requires interference between the CKM-leading and CKM-suppressed terms in eqs.~\eqref{eq:A-Dpipi}, \eqref{eq:A-DKK}. 
Neglecting tiny ${\cal O}(|\lambda_{b}/\lambda_{s,d}|^2)$ corrections, the asymmetries in the $\pi^+ \pi^-$ and $K^+ K^-$ final states read
\begin{align}
    a_{\rm CP}^{\rm dir} (D^0 \to \pi^+ \pi^-) &= 2 \left|\frac{\lambda_b}{\lambda_d} \right| \sin \gamma \left|\frac{{\cal P}_{\pi \pi}}{ {\cal A}_{\pi \pi}} \right| \sin \phi_{\pi \pi}\,,
    \label{eq:acp_pipi}
    \\[2mm]
    a_{\rm CP}^{\rm dir} (D^0 \to K^+ K^-) &= - 2 \left|\frac{\lambda_b}{\lambda_s} \right| \sin \gamma \left|\frac{{\cal P}_{KK}}{ {\cal A}_{K K}} \right| \sin \phi_{KK}\,,
    \label{eq:acp_KK}
\end{align}
where $\phi_{\pi \pi}$ and $\phi_{KK}$ denote the relative strong phases between ${\cal P}_{\pi \pi}$ and ${\cal A}_{\pi \pi}$, and between ${\cal P}_{KK}$ and ${\cal A}_{KK}$, respectively. Moreover, $\gamma \equiv - \arg (\lambda_b/\lambda_s)$ is the CKM angle.
The above equations show that
in the exact $U$-spin limit of QCD, the two direct CP asymmetries become equal in magnitude and opposite in sign, up to small differences of the order of $1- |\lambda_d/\lambda_s| = {\cal O}(10^{-4})$ originating from the CKM factors. 

The difference of the two asymmetries in the $K^+K^-$ and $\pi^+ \pi^-$ modes, $\Delta a_{\rm CP}^{\rm dir} \equiv a_{\rm CP}^{\rm dir} (D^0 \to K^+ K^-) - a_{\rm CP}^{\rm dir} ( D^0 \to \pi^+ \pi^-)$ is then given by
\begin{equation}
    \Delta a_{\rm CP}^{\rm dir} = - 2 \left|\frac{\lambda_b}{\lambda_s} \right| \sin \gamma\left(  \left|\frac{{\cal P}_{KK}}{ {\cal A}_{K K}} \right| \sin \phi_{KK} + 
 \left|\frac{{\cal P}_{\pi \pi}}{ {\cal A}_{\pi \pi}} \right| \sin \phi_{\pi \pi} \right)\,,
 \label{eq:Delta_acp}
\end{equation}
where we have used $|\lambda_d| \simeq |\lambda_s|$.

In the SM, the amplitudes ${\cal A}_{\pi\pi}, {\cal A}_{KK}$ and ${\cal P}_{\pi\pi}, {\cal P}_{KK}$ in eqs.~\eqref{eq:A-Dpipi}, \eqref{eq:A-DKK} receive contributions from several topologies, associated with the different possible contractions of the quark fields in the effective operators. These include tree, exchange, and penguin contractions, see e.g.~\cite{Lenz:2023rlq} for some illustrative diagrams. A precise determination of the hadronic quantities in eqs.~\eqref{eq:acp_pipi}, \eqref{eq:acp_KK} is therefore a remarkably challenging task. 

An estimate of the factorisable contributions from the tree-level contractions of the current-current operators at LO was obtained in~\cite{Lenz:2023rlq} using LCSR. The results were found to be consistent with naive factorisation and yielded predictions for the corresponding branching ratios in good agreement with experimental measurements, albeit with very large uncertainties. The LCSR framework had previously been employed in~\cite{Khodjamirian:2017zdu} to estimate the size of the penguin contractions of the current-current operators, which contribute to the first term of the CKM-subleading amplitudes ${\cal P}_{\pi\pi}$ and $ {\cal P}_{KK}$ in eq.~\eqref{eq:P}. 

In both analyses, the contributions of the QCD penguin operators $Q_k$ in eq.~\eqref{eq:Heff} were neglected due to the smallness of their Wilson coefficients. However, the $(V-A)\otimes(V+A)$ Dirac structure of the QCD penguin operators $Q_5$ and $Q_6$ gives rise to a chiral enhancement relative to the remaining $(V-A)\otimes(V-A)$ operators. This becomes manifest after Fierz transforming $Q_5$ and $Q_6$ as
\begin{equation}
Q_5 = -  2\sum_{q=u,d,s} \left(\bar u^i  q^j \right)_{S+P} \left(\bar q^j   c^i \right)_{S-P}\,, 
\qquad
Q_6 = -2 \sum_{q=u,d,s} \left(\bar u^i  q^i \right)_{S+P}  \left(\bar q^j   c^j \right)_{S-P}\,,
\label{eq:Penguin_Fierz}
\end{equation}
with $(\bar q_1 q_2)_{S\pm P}\equiv \bar q_1(1\pm\gamma_5)q_2$. Focusing on the $\pi^+ \pi^-$ mode, with analogous expressions for the $K^+ K^-$ channel, and singling out the $q = d $ component of the QCD penguin operators, leads to the following relations valid in the naive factorisation limit:
\begin{equation}
\begin{cases}
\langle  \pi^+ \pi^- | Q_4^d |D^0 \rangle  = \langle  \pi^+ \pi^- | Q_1^d |D^0 \rangle\,,\\[2mm]
\langle  \pi^+ \pi^- | Q_6^d |D^0 \rangle  =  \displaystyle{\frac{2 \mu_\pi}{m_c}}\langle  \pi^+ \pi^- | Q_1^d |D^0 \rangle\,, 
\end{cases} 
\qquad \mbox{(factorisation)}
\end{equation}
where $\mu_\pi \equiv m_\pi^2/(m_u + m_d)$ is the chirally-enhanced parameter.
The corresponding relations for the remaining operators $Q_3^d$ and $Q_5^d$ differ only by an overall colour factor.  We recall that, in naive factorisation,
\begin{equation}
    \langle  \pi^+ \pi^- | Q_1^d |D^0 \rangle  = i f_\pi (m_D^2 - m_\pi^2) f_0^{D\pi} (m_\pi^2)\,, \qquad \mbox{(factorisation)}
\end{equation}
where $f_\pi$ is the pion decay constant and $f_0^{D \pi}$ the scalar $D \to \pi$ form factor. 
The factorised matrix element of $Q_6^d$ is therefore enhanced with respect to that of the current-current operator $Q_1^d$.
For instance, using $\mu_\pi = 2.5$ GeV~\cite{Khodjamirian:2017fxg} and $m_c = 1.27 $ GeV~\cite{ParticleDataGroup:2026aaa} in the $\overline{\rm MS}$ scheme, gives approximately a factor four enhancement. 

Moreover, as pointed out in the study of $B \to \pi \pi$ decays~\cite{Khodjamirian:2005wn}, there is an annihilation-type factorisable contribution originating from the $q = u$ component of the operators $Q_5$ and $Q_6$ in eq.~\eqref{eq:Penguin_Fierz}. 
For the colour-singlet combination $Q_6^u = -2 \left(\bar u (1+ \gamma_5) u \right)  \left(\bar u   (1-\gamma_5) c \right)$ the corresponding matrix element in factorisation becomes proportional to the $D$-meson decay constant $f_D$ and the pion scalar form factor $F^S_\pi$,~\footnote{This is defined as $\langle \pi^+ (p_1) \pi^- (p_2) | \bar u u |0 \rangle = F_\pi^S \big( (p_1+p_2)^2\big)$.} yielding
\begin{equation}
    \frac{\langle \pi^+ \pi^-| Q_6^u|D^0\rangle}{\langle \pi^+ \pi^-| Q_1^d|D^0\rangle} = - \frac{ 2 m_D^2 }{m_c (m_D^2 - m_\pi^2)} \frac{f_D}{f_\pi} \frac{ F_\pi^S(m_D^2)}{f_0^{D\pi} (m_\pi^2)}\,. \qquad \mbox{(factorisation)}
\end{equation} 
For $B \to \pi\pi$ decays, the corresponding effect was found to be sizeable~\cite{Khodjamirian:2005wn}. A dedicated study to assess its relevance in $D^0$ decays is therefore of particular interest.

Motivated by these observations, we extend the analyses of~\cite{Khodjamirian:2017zdu,Lenz:2023rlq} by estimating, within LCSR, the matrix elements of the QCD penguin operators entering ${\cal P}$ in eq.~\eqref{eq:P}.
%%%%%%%%%%%%%%%%%%%%%%%%%%%%%%%%%%%%%%%%%%%%%%%%%
%%%%%%%%%%%%%%%%%%%%%%%%%%%%%%%%%%%%%%%%%%%%%%%%%
%%%%%%%%%%%%%%%%%%%%%%%%%%%%%%%%%%%%%%%%%%%%%%%%%
\section{The hadronic matrix elements from LCSR}
\label{sec:LCSR}
In this section, we present the LCSR calculation of the tree-level matrix elements of the QCD penguin operators $Q_3, \ldots, Q_6,$ in eq.~\eqref{eq:Heff}. For consistency, and owing to differences between the framework adopted here and that used in the previous analysis~\cite{Lenz:2023rlq}, we also revisit the calculation of the matrix elements of the current-current operators $Q^q_1$ and $Q_2^q$.

For definiteness, we focus on $D^0 \to \pi^+ \pi^-$; the extension to $D^0 \to K^+ K^-$ is straightforward, and the relevant differences will be pointed out where necessary. Throughout this work, we neglect explicit light-quark mass effects by setting $m_u = m_d = m_s =0$, while retaining the chirally enhanced parameters $\mu_\pi$ and $\mu_K$. Therefore, we account for $SU(3)_f$-breaking effects only through the hadronic parameters of the kaon, including its mass, decay constant, and the Gegenbauer moments entering the kaon distribution amplitudes.

Our analysis is based on the framework developed for the $B \to \pi\pi$ system in~\cite{Khodjamirian:2000mi,  Khodjamirian:2005wn} (see also~\cite{Khodjamirian:2003xk}), which was more recently adapted to the modes
$D^0 \to \pi^+ \pi^-$ and $D^0 \to K^+ K^-$ in~\cite{Lenz:2023rlq}. 
Specifically, we introduce the following three-point correlation function
\begin{equation}
F^{O}_{\mu} (p, q, k) = i^2 \int d^4 x \, e^{- i p \cdot x} \int d^4 y
\, e^{i (q - k) \cdot y} \,
\langle \pi^- (p - q)| {\rm T} \! \left\{  j^{D}_5 (x), O(0), j^{\pi}_\mu (y)\right\} |0 \rangle\,,
\label{eq:Correlator-1}
\end{equation}  
where $O$ denotes one of the effective four-quark operators defined in eqs.~\eqref{eq:Q1}, \eqref{eq:QC_P}. The pion in the external state is on-shell, $(p-q)^2 = m_{\pi}^2 = 0$, while the interpolating currents of the $D^0$ and $\pi^+$ mesons are chosen as $j^D_5 = i  m_c \, \bar c \gamma_5 u $ and $j^{\pi}_\mu =  \bar d \gamma_\mu \gamma_5 u$, respectively, with corresponding off-shell four-momenta $p$ and $ q - k $. 
Following~\cite{Khodjamirian:2000mi}, we have introduced in the correlation function an auxiliary four-momentum $k$, to ensure that the dispersion relation in the $D$-meson channel is not contaminated by contributions from unphysical or ``parasitic'' singularities below the physical $D$-meson pole. Thus, in this configuration, the auxiliary momentum $k$ flows through the weak vertex. As shown below, the  dependence on $k$ disappears upon taking the physical limit in the ground-state contribution, leading to the desired hadronic matrix element.~\footnote{A similar approach, in which the weak Hamiltonian carries momentum, was proposed in~\cite{Buchler:2001nm} in the context of $K \to \pi \pi$ decays to account for the effects of final-state interactions.} 
We note that the analysis in~\cite{Lenz:2023rlq} for the current-current operators employed a different three-point correlation function without the auxiliary momentum $k$. 
The differences between these two approaches will  be discussed later on.

Taking into account the Lorentz structure of $F^O_\mu (p, q, k)$, we decompose the correlation function as
\begin{equation}
F^O_\mu(p,q,k) =  
 (q- k)_\mu F^O  +
 p_\mu F^O_1 + q_\mu F^O_2 + \epsilon_{\mu \nu \rho \sigma} p^\nu q^\rho k^\sigma F^O_3\,,
\label{eq:Fq-Fp}
\end{equation}
where $F^O, \ldots, F_3^O$ are scalar functions depending on the six independent Lorentz invariants that can be constructed from the three four-momenta $p,k,q$. We choose these invariants as 
\begin{equation}
p^2\,, \, q^2\,, \, k^2\,, \, (q-k)^2\,, \,  (p-k)^2\,, (p-q)^2\,,
\end{equation}
where $(p-q)^2$ is fixed by the pion on-shell condition, 
and, following~\cite{Khodjamirian:2000mi}, we further set $k^2 = q^2 = 0$, since these invariants do not correspond to physical dispersion variables and may therefore be chosen conveniently.
With the introduction of the auxiliary momentum $k$, the variable $p^2$
entering the dispersion relation in the $D$-meson channel becomes independent of $P^2 \equiv (p-k)^2$, corresponding to the invariant mass squared of the configuration after the weak decay. 
In the following, we focus on the invariant amplitude $F^O$, i.e., the coefficient of $(q-k)_\mu$, since this Lorentz structure is the only one relevant for the construction of the hadronic dispersion relations. 

For sufficiently large and spacelike invariant momentum variables, the correlation function in eq.~\eqref{eq:Correlator-1} can be computed in QCD. To this end, we consider the kinematical region
\begin{equation}
- P^2 \gg \Lambda^2\,, \quad - p^2 \gg \Lambda^2\,, \quad - (q-k)^2 \gg \Lambda^2\,,
\label{eq:kinematics}
\end{equation}
with $\Lambda$ denoting a hadronic scale of order a few hundred MeV. In this regime, the light-cone expansion applies and the dominant contribution to the double integral in eq.~\eqref{eq:Correlator-1} originates from the light-cone region $x^2 \sim y^2 \sim (x-y)^2 \sim 0$, see e.g.~\cite{Khodjamirian:2000mi, Khodjamirian:2005wn, Khodjamirian:2020btr}.
With the kinematics fixed above, 
the correlation function in eq.~\eqref{eq:Correlator-1} can be computed in terms of a light-cone operator-product expansion (LC-OPE) as a convolution of hard scattering kernels with the corresponding pion light-cone distribution amplitudes~(LCDAs) of growing twist. 
The resulting OPE admits the same Lorentz decomposition as in eq.~\eqref{eq:Fq-Fp} and provides an analytic expression for the invariant amplitude $F^O\big((q-k)^2, p^2, P^2\big)|_{\rm OPE}$. The explicit calculation of the LC-OPE is presented in the next section. 

The LC-OPE result must then be matched to hadronic dispersion relations, constructed in the $p^2$ and $(q-k)^2$ channels, in order to isolate the desired hadronic matrix element. The details of these derivations can be found in the original references~\cite{Khodjamirian:2000mi,Khodjamirian:2005wn}; here, we restrict ourselves to outlining the main steps relevant for our analysis.

We start by deriving the hadronic dispersion relation in the $(q-k)^2$ channel, keeping $p^2$ and $P^2$ fixed. 
By inserting a complete set of hadronic states with the quantum numbers of the pion into eq.~\eqref{eq:Correlator-1} and isolating the ground-state contribution from the higher resonances and continuum states, we obtain
\begin{equation}
    F^O\big((q-k)^2, p^2, P^2 \big) =  \frac{i f_\pi\, \Pi_{\pi \pi}^O (p^2, P^2)}{m_\pi^2 - (q-k)^2}  + \int_{s_h^{(\pi)}}^\infty ds^\prime \, \frac{\rho_h^{(\pi)} ( s^\prime, p^2, P^2)}{s^\prime - (q-k)^2}\,, 
    \label{eq:DR1}
\end{equation}
where $s_h^{(\pi)}$ is the hadronic continuum threshold, $f_\pi$ the pion decay constant, and  $\Pi_{\pi \pi}^O$ is the two-point correlation function 
\begin{equation}
    \Pi_{\pi \pi}^O (p^2, P^2) = i \int d^4 x \, e^{- i p \cdot x}
\langle \pi^- (p - q) \pi^+ (q-k)| {\rm T} \! \left\{j^{D}_5 (x), O(0) \right\} |0 \rangle \,.
\label{eq:DR2}
\end{equation}
This correlation function is understood as an analytic function of the external momenta. By crossing symmetry, it can equivalently be interpreted as a pion scattering correlation function at large and spacelike momentum transfer  squared $P^2$~\cite{Khodjamirian:2000mi}. 

Next, we match the hadronic dispersion relation in eq.~\eqref{eq:DR1} to the dispersive representation of the OPE result in the complex variable $s^\prime = (q-k)^2$, while keeping the remaining invariants $p^2$ and $P^2$ large and negative. Hence, invoking quark-hadron duality~(QHD), we approximate the contribution of higher resonances and continuum states on the r.h.s.\ of eq.~\eqref{eq:DR1} by the corresponding dispersive integral over the OPE spectral density above an effective threshold $s_0^\pi$, i.e.
\begin{equation}
    \int_{s_h^{(\pi)}}^\infty ds^\prime \, \frac{\rho_h^{(\pi)} ( s^\prime, p^2, P^2)}{s^\prime - (q-k)^2} \overset{\rm QHD}{=} 
    \frac{1}{\pi} \int_{s_0^{\pi}}^\infty d s^\prime \, \frac{{\rm Im}_{s^\prime} \,F^O\big(s^\prime, p^2, P^2\big)_{\rm OPE}}{s^\prime - (q-k)^2}\,.
    \label{eq:QHD}
\end{equation}
Here, $s_0^\pi$ is a sum-rule parameter to be further determined and should therefore be distinguished from the physical hadronic continuum threshold $s_h^{(\pi)}$.
To suppress the contribution of higher resonances and continuum states, and thereby reduce the sensitivity to the QHD approximation, we perform a Borel transformation with respect to the variable $(q-k)^2$, under which 
\begin{equation}
\frac{1}{s^\prime - (q-k)^2} \to \frac{1}{M^{\prime 2}} e^{- s^\prime/M^{\prime 2}}\,,
\end{equation}
where $M^{\prime 2}$ is the Borel parameter.
Having derived a dispersion relation in the pion channel and obtained an analytic expression for $\Pi_{\pi \pi}^O (p^2, P^2)$, we can analytically continue this correlation function to the timelike $P^2$ region, evaluating it at the physical point $P^2 = m_D^2$, before proceeding to the derivation of the dispersion relation in the $p^2$ channel. To this end, we follow the same steps as before and insert into eq.~\eqref{eq:DR2} a complete set of states with the $D^0$-meson quantum numbers. After isolating the ground state contribution, we obtain
\begin{equation}
    \Pi_{\pi \pi}^O (p^2, m_D^2) = \frac{f_D m_D^2 \, \langle \pi^- (p - q) \pi^+ (q)|    O |D^0 (p) \rangle }{m_D^2 - p^2} + \int_{s_h^{(D)}}^\infty d s \, \frac{\rho_h^{(D)}(s)}{s- p^2}\,.
    \label{eq:DR3}
\end{equation}
Note that we have recovered the desired physical matrix element $\langle\pi^+\pi^-|O|D^0\rangle$, since in the ground state contribution the dependence on the auxiliary four-momentum $k$ vanishes. At the $D$-meson pole one has $p^2=m_D^2$, while the final-state invariant satisfies $P^2=(p-k)^2=m_D^2$. Together with $k^2=0$, these conditions imply $p\cdot k=0$, and hence $k^\mu=0$ for a physical timelike $D$-meson momentum. The dependence on the four-momentum $k$, however, remains in the continuum contribution described by the spectral density $\rho_h^{(D)}$~\cite{Khodjamirian:2000mi}. 

Expressing the OPE result for the two-point correlation function $\Pi_{\pi \pi}^O (p^2, m_D^2)$ in terms of a dispersive integral in the complex variable $s = p^2$ we can use again QHD to approximate the continuum contribution in eq.~\eqref{eq:DR3} with the integral of the OPE spectral density above an effective threshold $s_0^D$. Finally, performing a Borel transformation in the variable $p^2$, with Borel parameter $M^2$, leads to the following result for the hadronic matrix element
\begin{equation}
    i \langle \pi^+ \pi^- |O| D^0 \rangle = \frac{e^{m_{\pi}^2/M^{\prime 2}} e^{m_{D}^2/M^2}}{\pi^2 f_\pi f_D \, m_D^2} \int\limits_{m_c^2}^{s_0^D} d s   \int\limits_{0}^{s_0^\pi} d s^\prime \! \, e^{-s/M^2} \, e^{-s^\prime/M^{\prime 2}}\,  {\rm Im}_s {\rm Im}_{s^\prime}  F^O \big(s^\prime, s, m_D^2 \big)_{\rm OPE} \,.
    \label{eq:SR}
\end{equation}
The corresponding result for the mode $D^0 \to K^+ K^-$ is obtained replacing $f_\pi \to f_K$, $m_\pi \to m_K$ and $s_0^\pi \to s_0^K$. The calculation of the OPE is discussed in the next section.
%%%%%%%%%%%%%%%%%%%%%%%%%%%%%%%%%%%%%%%%%%%%%%
%%%%%%%%%%%%%%%%%%%%%%%%%%%%%%%%%%%%%%%%%%%%%%
%%%%%%%%%%%%%%%%%%%%%%%%%%%%%%%%%%%%%%%%%%%%%%
\section{Derivation of the LC-OPE}
\label{sec:OPE}
In this section, we derive the LC-OPE of the three-point correlation function in eq.~\eqref{eq:Correlator-1} at leading order in $\alpha_s$, for all four-quark operators defined in eqs.~\eqref{eq:Q1}, \eqref{eq:QC_P}. The result is expressed in terms of short-distance functions convoluted with non-local vacuum-to-pion or vacuum-to-kaon matrix elements, which are then parametrised by LCDAs, as well as by products of LCDAs and quark condensates.  

For clarity, below we restrict the discussion to four-quark operators of the form
\begin{equation}
    \big( \bar u \Gamma_1^O c) \big( \bar q \Gamma_2^O q\big)\,, \qquad q = u,d,s\,,
    \label{eq:operator}
\end{equation}
where the colour indices are implicitly summed over and $\Gamma_{1}^O$ and $\Gamma_{2}^O$ denote the Dirac structures of the $(\bar u c)$ and $(\bar q q)$ bilinears, respectively, for the specific operator $O$.~\footnote{The current-current operators in eq.~\eqref{eq:Q1} are brought to this form using a Fierz transformation.} Hence, the discussion explicitly applies to the operators $O = \{ Q_2^q, Q_3$, $Q_5\}$. 
The remaining operators are related to these by means of the identity 
\begin{equation}
    t^a_{ij} t^a_{lm } = \frac12 \left( \delta_{im} \delta_{jl} - \frac{1}{N_c} \delta_{ij} \delta_{lm} \right)\,,
    \label{eq:Fierz_id}
\end{equation}
where $t^a_{ij}$ are the generators of the fundamental representation of $SU(3)_c$ and $N_c = 3$ is the number of colours, see appendix~\ref{sec:app1p}.
%%%%%%%%%%%%%%%%%%%%%%%%%%%%%%%%%%%%%%%%%%%%%%%%%
%%%%%%%%%%%%%%%%%%%%%%%%%%%%%%%%%%%%%%%%%%%%%%%%%
\subsection{Contributions from quark loops}
The first class of contributions to the correlation function in eq.~\eqref{eq:Correlator-1} is obtained by contracting all quark fields in the time-ordered product except for the external $\bar d u$ fields. The resulting quark propagators are evaluated near the light cone using their expansion in the external background gluon field, see e.g.~\cite{Belyaev:1994zk}. At the level of accuracy considered here, however, it suffices to retain only the leading term, corresponding to zeroth order in the external gluon field. The quark propagators therefore reduce to the free propagators.

\begin{figure}[t]
    \centering

    \begin{subfigure}[b]{0.33\textwidth}
        \centering
        \includegraphics[width=\textwidth]{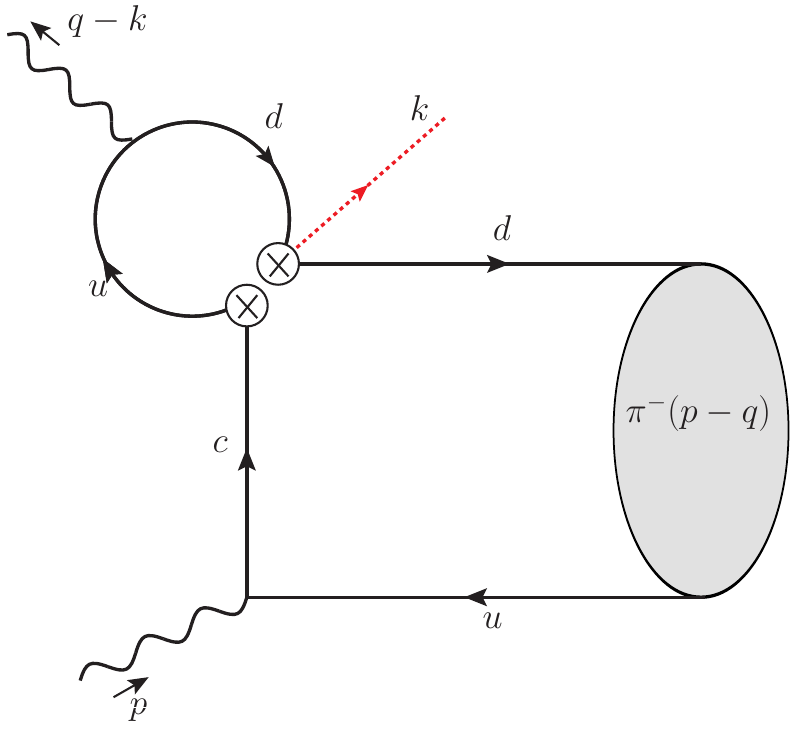}
        \caption{${\rm P}_{\rm I}$}
        \label{fig:T1}
    \end{subfigure}
    \qquad
    \qquad
    \begin{subfigure}[b]{0.33\textwidth}
        \centering
        \includegraphics[width=\textwidth]{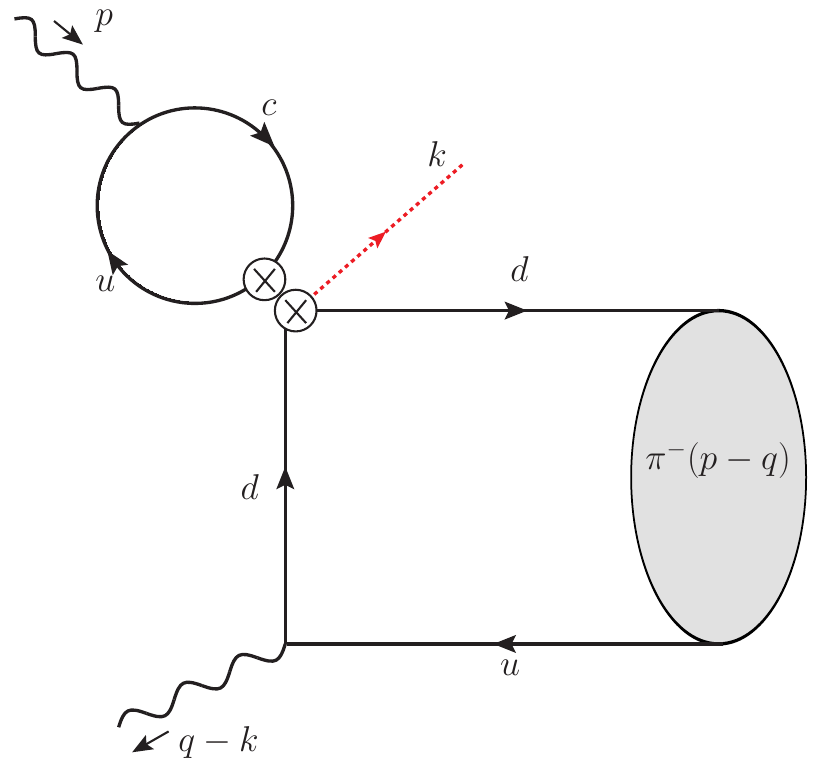}
        \caption{${\rm P}_{\rm II}$}
        \label{fig:T2}
    \end{subfigure}
     \vspace{1em}
    \begin{subfigure}[b]{0.35\textwidth}
        \centering
        \includegraphics[width=\textwidth]{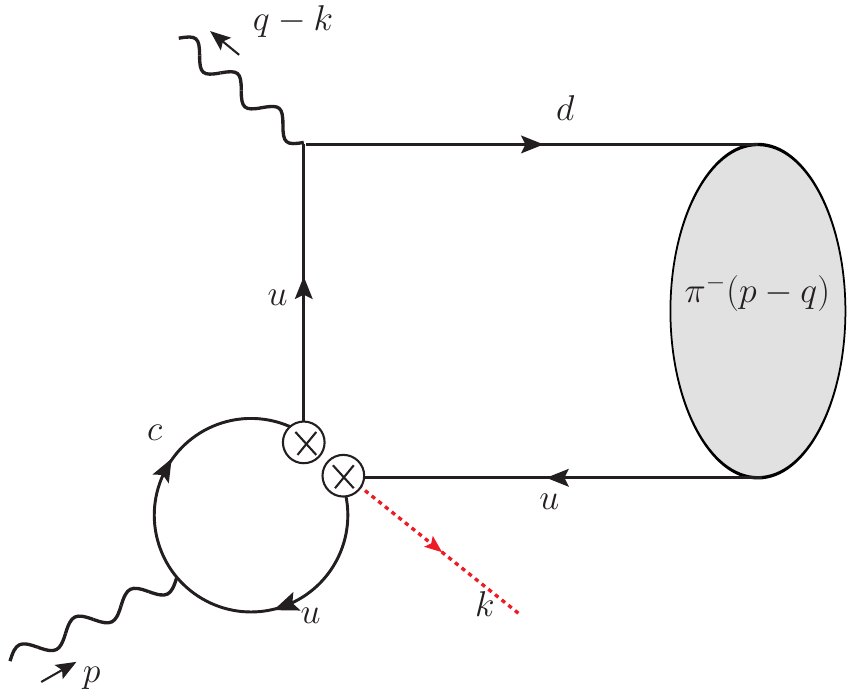}
        \caption{${\rm P}_{\rm III}$}
        \label{fig:T3}
    \end{subfigure}
    \qquad
    \quad
    \begin{subfigure}[b]{0.35\textwidth}
        \centering
        \includegraphics[width=\textwidth]{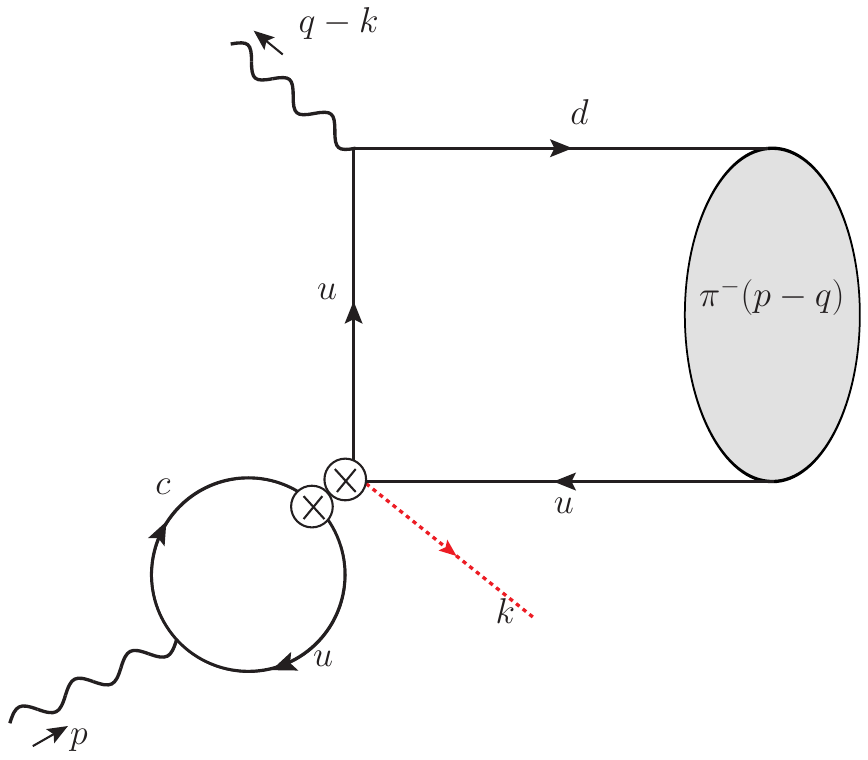}
        \caption{${\rm P}_{\rm IV}$}
        \label{fig:T4}
    \end{subfigure}

    \caption{Quark-loop contractions contributing at leading order in $\alpha_s$ to the correlation function in eq.~\eqref{eq:Correlator-1}. The auxiliary four-momentum $k$ is indicated by the dotted red line, while the crossed circles denote the insertion of one of the effective four-quark operators in the effective Hamiltonian. }
    \label{fig:Topologies}
\end{figure}

Depending on the flavour structure of the operator inserted in eq.~\eqref{eq:Correlator-1}, up to four distinct quark-field contractions contribute. 
In the case of the current-current operator $Q_2^q$, where $q$ is a down-type quark, two connected Wick contractions are possible, as shown in the panels (a) and (b) of figure~\ref{fig:Topologies} for the decay $D^0 \to \pi^+ \pi^-$. The corresponding diagrams for $D^0 \to K^+ K^-$ are obtained by replacing $d \to s$.

However, for the penguin operators, $Q_{3,5} = \sum_{q=u,d,s} Q_{3,5}^q$, besides the contributions with $q=d,s$, two additional contractions arise from the $q=u$ component. These are shown in the panels (c) and (d) of figure~\ref{fig:Topologies}, again for the $D^0 \to \pi^+ \pi^-$ channel.

The quark-loop contribution to the correlation function can thus be decomposed as
\begin{equation}
    F_\mu^O (p,q,k)_{\rm QL} = \int d^4 x \, e^{- i p \cdot x} \int d^4 y
\, e^{i (q-k) \cdot y} \, \big( \mbox{P}^O_{{\rm I}} + \mbox{P}_{{\rm II}}^O + \mbox{P}_{{\rm III}}^O + \mbox{P}_{{\rm IV}}^O \big)_\mu (p,q; x,y)\,,
\end{equation}
where
\begin{equation}
{\rm P}^{O}_{ {\rm I} \mu}  = - m_c \,
\langle \pi^- (p - q)|\biggl[\bar d(0) \Gamma_2^O S_0^{(d)}(-y) 
 \gamma_\mu \gamma_5 S_0^{(u)}(y) \Gamma_1^O S_0^{(c)}(-x) \gamma_5 u(x)\biggr]|0\rangle \,,
\label{eq:T1}
\end{equation}

\begin{equation}
{\rm P}^{O}_{{\rm II}\mu}  = m_c N_c \, 
\langle \pi^- (p - q)|\biggl[\bar d(0) \Gamma_2^O S_0^{(d)}(-y) 
 \gamma_\mu \gamma_5 u(y) \biggr]|0\rangle \, {\rm Tr} \bigg[\Gamma_1^O S_0^{(c)}(-x) \gamma_5 S_0^{(u)}(x) \bigg]  \,,
\label{eq:T2}
\end{equation}

\begin{equation}
{\rm P}^{O}_{{\rm III}\mu} = - m_c  \, 
\langle \pi^- (p - q)|\biggl[\bar d(y) \gamma_\mu \gamma_5 S_0^{(u)}(y) \Gamma_1^O 
S_0^{(c)}(-x) \gamma_5 S_0^{(u)}(x) \Gamma_2^O u(0)\biggr]|0\rangle \,,
\label{eq:T3}
\end{equation}

\begin{equation}
{\rm P}^{O}_{{\rm IV}\mu}  = m_c N_{c}\,
\langle \pi^- (p - q)|\biggl[\bar{d}(y) \gamma_\mu \gamma_5  S_0^{(u)}(y) \Gamma_2^O
 u(0) \biggr]|0\rangle \, {\rm Tr} \biggl[\Gamma_1^O S_0^{(c)}(-x) \gamma_5 S_0^{(u)}(x) \biggr]  \,.
\label{eq:T4}
\end{equation}
Here, $S_{0}(x-y)$ is the free-quark propagator, with the corresponding flavour indicated by the superscript. 
As already discussed above, the contributions ${\rm P}^{O}_{{\rm III}} $ and ${\rm P}^{O}_{{\rm IV}}$ are absent when considering the current-current operator, i.e.\ for $O = Q_2^q$.

Note that the different contractions have also a direct interpretation in terms of the underlying topologies. In the case of ${\rm P}^{O}_{{\rm I}} $, the $\bar u d$ component of the effective four-quark operator is contracted with the interpolating current of the light meson, here the pion, forming a quark loop. Diagrammatically, this corresponds to the tree-level topology shown in panel (a) of figure~\ref{fig:Topologies}. 
In the remaining three contributions, it is the $\bar c u$ component of the effective operator that is contracted with the interpolating current of the $D$ meson, forming the corresponding quark loop. These contributions can be identified with annihilation-type topologies, shown in panels (b), (c), and (d) of figure~\ref{fig:Topologies}.

For each of the contributions in eqs.~\eqref{eq:T1} - \eqref{eq:T4}, the non-local vacuum-to-pion quark-antiquark matrix elements, with the quark fields separated along either the $x$ or $y$ coordinate, are parametrised in terms of pion LCDAs up to twist four. 
The general light-cone expansion of the pion two-particle matrix elements can be found e.g.\ in~\cite{Duplancic:2008ix} (see~\cite{Khodjamirian:2009ys} for the corresponding kaon case), and reads
\begin{align}
 \langle \pi^- (p)| \bar d^i_{\alpha} (x_1) u_{\beta}^j (x_2) |0 \rangle_{x_i^2 \sim 0} & =  i f_\pi \, \frac{ \delta^{ij}}{12} \int\limits_0^1 d u \, e^{i u p \cdot x_1 + i \bar u p \cdot x_2} \Biggl(  \phi_{2\pi} (u) \slashed p \gamma_5
 \label{eq:2pME}
\\
&  \hspace{-10mm} - \, \mu_\pi \phi_{3 \pi}^p (u) \gamma_5  + \frac{1}{6}\mu_\pi \phi_{3 \pi}^\sigma (u)   p^{\mu} (x_1 - x_2)^\nu \sigma_{\mu \nu} \gamma_5 
\nonumber \\
&  \hspace{-10mm} + \, \frac{1}{16} \phi_{4 \pi}(u) (x_1 - x_2)^2  \slashed p \gamma_5 - \frac{i}{2} (\slashed x_1 - \slashed x_2) \gamma_5 \int\limits_{0}^u d v \psi_{4 \pi} (v) \Biggr)_{\beta \alpha}\,,
\nonumber
\end{align}
where $p$ here indicates the pion momentum, $\sigma_{\mu \nu} = (i/2)[\gamma_\mu, \gamma_\nu]$, $\alpha$ and $\beta$ denote spinor indices, $\phi_{2\pi}, \phi_{3\pi}^p, \phi_{3 \pi}^\sigma, \phi_{4\pi}$, and $\psi_{4\pi}$ are the pion LCDAs of twist two, three, and four, respectively, $\bar u = 1-u$, and $\mu_\pi$ is the chirally-enhanced parameter. 

Substituting the above expression into eqs.~\eqref{eq:T1} - \eqref{eq:T4} reduces the calculation of the correlation function to the evaluation of one-loop integrals multiplying the corresponding pion LCDAs. Depending on the topology, the loop integration originates from either the $x$- or the $y$-integration. We evaluate these one-loop integrals in momentum space using dimensional regularisation in $d=4-\epsilon$ dimensions together with the anticommuting prescription for $\gamma_5$. This does not lead to any ambiguities, since all non-vanishing Dirac traces contain an even number of $\gamma_5$ matrices. From the loop integrals, we retain only the finite part depending on $p^2$ or $(q-k)^2$, as this is the only contribution relevant for constructing the dispersion relations at leading order in the strong coupling considered here. The remaining $y$- or $x$-integration is conveniently performed directly in coordinate space, using the representation of the free-quark propagator in terms of Bessel functions together with the master integrals collected, for example, in the appendix of~\cite{Piscopo:2023opf}.

Having described the general method to compute eqs.~\eqref{eq:T1} - \eqref{eq:T4}, we now turn to discuss the individual contractions, distinguishing the tree-level topology in eq.~\eqref{eq:T1} from the remaining annihilation contributions in  eqs.~\eqref{eq:T2} - \eqref{eq:T4}.
%%%%%%%%%%%%%%%%%%%%%%%%%%%%%%%%%%%%%%%%%%%%%%%%%
%%%%%%%%%%%%%%%%%%%%%%%%%%%%%%%%%%%%%%%%%%%%%%%%%
\subsubsection{The tree-level topology
}
Evaluating the tree-level contraction in eq.~\eqref{eq:T1} yields the following LC-OPE for the invariant amplitude $F^O$:
\begin{align}
     F_{{\rm I}}^{O}  \big((q-k)^2,p^2, P^2 \big)_{\rm QL} = f_\pi m_c    \int_0^1   d u & \sum_\phi \phi(u) \sum_{n = 1}^3 \frac{c_{\phi,n, {\rm I} }^{O} \big( u, P^2, (q-k)^2\big)}{\big( \tilde s_{\rm I}   (u)- p^2 - i \varepsilon\big)^n} 
     \nonumber \\[1mm]
     & \times \ln\left( -\frac{(q-k)^2 + i \varepsilon}{\mu^2}  \right)\,,
     \label{eq:FI_OPE}
\end{align}
where the first sum runs over the pion LCDAs of twist two, three, and four, i.e., $\phi=\phi_{2\pi},\ldots,\psi_{4\pi}$; $\mu$ denotes the renormalisation scale and we have defined
\begin{equation}
\tilde s_{\rm I}(u)\equiv \frac{m_c^2 + u \bar u m_\pi^2}{u} \,.
\label{eq:stilde_I}
\end{equation} 
Note that we keep the pion mass explicit to facilitate comparison with the kaon case, which is obtained by replacing $m_\pi \to m_K$ together with the corresponding pion LCDAs by their kaon counterparts. In the numerical analysis, however, we set $m_\pi=0$. 

The explicit expressions for the coefficient functions $c_{\phi,n, {\rm I}}^O$ in eq.~\eqref{eq:FI_OPE} are collected in appendix~\ref{sec:app2}. 
These coefficients are non-vanishing and identical for $O = Q_2^d, Q_3$, whereas they vanish for $O = Q_5$. This difference originates from the $(V-A) \otimes (V+A)$ Dirac structure of the operator $Q_5$, which makes the contraction in eq.~\eqref{eq:T1} proportional to the light-quark masses and hence chirally suppressed. Consequently, this contribution vanishes in the limit $m_{d,u} = 0$.~\footnote{This contribution remains zero also after including soft-gluon corrections, as these do not alter the Dirac structure of the correlation function.} This applies to the mode $D^0 \to K^+ K^-$ too, since we work in the limit $m_s=0$. However, even if we kept a non-zero strange-quark mass, this contribution would remain strongly suppressed, being proportional to $m_s$.

From the expression in eq.~\eqref{eq:FI_OPE}, the imaginary part entering the final sum rule in eq.~\eqref{eq:SR} is obtained using the results collected in appendix~\ref{sec:app1}. The invariant amplitude in eq.~\eqref{eq:FI_OPE} exhibits the expected analytic structure: as a function of $s'=(q-k)^2$, it has a branch cut along the positive real axis, $s'>0$, originating from the logarithm, while as a function of $s=p^2$, it develops a cut starting at $s = \tilde s_{\rm I}(u)$, with the lowest threshold given by $s=m_c^2$. The latter has support only on the integration region $u \geq u_{\rm min}$, where~\footnote{In the limit $m_\pi \to 0$, eq.~\eqref{eq:umin} reduces to $u_{min} = m_c^2/s_0^D$.}
\begin{equation}
u_{min} = \frac{1}{2} \sqrt{\frac{4 m_c^2 m_\pi^2+ m_\pi^4-2 m_\pi^2
   s_0^D + {s_0^{D}}^2}{m_\pi^4}}+\frac{m_\pi^2-s_0^D}{2 m_\pi^2}\,.
   \label{eq:umin}
\end{equation}
The dispersion integral over $s'$ can therefore be carried out analytically, after which we set $P^2=m_D^2$ before performing the remaining integrations over $s$ and $u$. 

It is instructive to compare the result in eq.~\eqref{eq:FI_OPE} with the one obtained without introducing the auxiliary four-momentum $k$ in the correlation function in eq.~\eqref{eq:Correlator-1}, as was done in~\cite{Lenz:2023rlq}. In that case, the invariant amplitude takes the form
\begin{equation}
     \hat F_{{\rm I}}^{O} \big(q^2,p^2 \big)|_{\rm QL} = f_\pi m_c \int_0^1 d u \sum_\phi \phi(u) \sum_{n = 1}^3 \frac{\hat c_{\phi,n, {\rm I}}^{O} \big( u,  q^2\big)}{\big( \hat{\tilde s}_{\rm I} (u, q^2) - p^2 - i \varepsilon\big)^n} \ln\left( -\frac{q^2 +  i \varepsilon}{\mu^2}\right)\,,
     \label{eq:FI_OPE_hat}
\end{equation}
where 
\begin{equation}
\hat{\tilde s}_{\rm I} (u, q^2)\equiv \frac{m_c^2 - \bar u q^2 + u \bar u m_\pi^2}{u} \,,
\label{eq:stilde_nok}
\end{equation}
in agreement with the expressions quoted in~\cite{Lenz:2023rlq} in the case of the current-current operators. Unlike eq.~\eqref{eq:FI_OPE}, the variable $q^2$ enters not only through the logarithm but also through the denominator via the function $\hat{\tilde s}_{\rm I}(u,q^2)$. Consequently, the singularities in the $s$- and $s'$-channels can, in principle, no longer be treated independently, making the construction of a dispersion relation in $s'$ less straightforward. However, the additional ``parasitic'' cut induced by $\hat{\tilde s}_{\rm I}(u,q^2)$ contributes only above $m_c^2$ and is therefore eliminated by the QHD approximation, as it lies outside the integration region bounded by $s_0^\pi$. This explains why the tree-level contribution could be evaluated in~\cite{Lenz:2023rlq} without the need to introduce the auxiliary momentum $k$. The numerical comparison between the two approaches is discussed in section~\ref{sec:Results}. As we shall see below, however, the situation changes significantly for the annihilation topologies.
%%%%%%%%%%%%%%%%%%%%%%%%%%%%%%%%%%%%%%%%%%%%%%%%%
%%%%%%%%%%%%%%%%%%%%%%%%%%%%%%%%%%%%%%%%%%%%%%%%%
\subsubsection{The annihilation topologies
}
The contribution of the annihilation diagrams in eqs.~\eqref{eq:T2} - \eqref{eq:T4} to the LC-OPE for the invariant amplitude $F^O$ can be written compactly as
\begin{align}
     F_{{\rm X}}^{O}  \big((q-k)^2,p^2, P^2 \big)_{\rm QL} = f_\pi m_c   \int_0^1   d u & \sum_\phi \phi(u)  \sum_{n = 1}^2  \frac{c_{\phi,n, {\rm X} }^{O} \big( u, P^2, p^2\big)}{\big( \tilde s^\prime_{\rm X}   (u, P^2) - (q-k)^2 - i \varepsilon\big)^n} 
     \nonumber \\[1mm]
    & \times \ln\left( \frac{m_c^2 - p^2 - i \varepsilon}{m_c^2}  \right)\,,
     \label{eq:F_II_IV_OPE}
\end{align}
where $X = \mbox{II, III, IV}$ and $\phi= \phi_{3 \pi}^p,\ldots, \psi_{4 \pi}$, i.e. the first sum runs only over the pion LCDAs of twists three and four. Moreover 
\begin{equation}
    \tilde s^\prime_{\rm II}   (u, P^2) = 
  \frac{- \bar u P^2 + u \bar u m_\pi^2}{u}\,, 
\end{equation}
while the corresponding expression for $X = \mbox{III, IV}$ is obtained with the replacement $u \to \bar u$. Analytical results for the coefficients $c_{\phi,n, {\rm X} }^{O}$ are collected in appendix~\ref{sec:app2}. 

As discussed above, the coefficients $c_{\phi,n,{\rm III}}^O$ and $c_{\phi,n,{\rm IV}}^O$ vanish in the case of current-current operator, i.e., for $O=Q_2^d$. Furthermore, for operators with  $(V-A)\otimes(V-A)$ Dirac structure, 
namely for $O=Q_2^d,Q_3$, only twist-four LCDAs contribute to eq.~\eqref{eq:F_II_IV_OPE} for all three annihilation topologies. The absence of leading-twist contributions, namely from twist-two and -three LCDAs, was already observed in~\cite{Khodjamirian:2005wn} in the case of current-current operators in $B\to \pi\pi$ decays. The latter work, however, did not include twist-four LCDAs. 

The situation is different for the remaining QCD penguin operator, i.e.\ for $O=Q_5$. 
While diagrams II and IV still depend only on subleading twist-four LCDAs, the contraction III receives contributions from twist-three LCDAs and is proportional to the chirally enhanced parameter $\mu_\pi$. This originates from the $(V-A)\otimes(V+A)$ Dirac structure of $Q_5$ and was first discussed in~\cite{Khodjamirian:2005wn} in the analysis of annihilation contributions to the $B\to\pi\pi$ system. This feature also emerges in naive factorisation, as discussed in section~\ref{sec:Heff}. 

Given its relevance for $Q_5$, below we discuss in detail the derivation of the dispersive representation of eq.~\eqref{eq:F_II_IV_OPE} for this operator, which closely follows the approach developed and discussed in~\cite{Khodjamirian:2005wn}.
The treatment of the remaining annihilation topologies proceeds analogously and is not repeated.~\footnote{We note that for all penguin operators, the contractions II and IV lead to contributions to the final sum rule which are related by an overall sign together with the replacement $u \to  \bar u$ in both the coefficient functions and the LCDAs. 
Since the pion LCDAs are symmetric under $u \to \bar u$, in particular $a_1^\pi=0$, the sum of these two contributions vanishes identically for $D^0\to\pi^+\pi^-$. This cancellation is absent for $D^0\to K^+K^-$, since the kaon LCDAs contain antisymmetric components, specifically those proportional to the first Gegenbauer moment $a_1^K\neq 0$, which change sign under $u \to \bar u$, leading to a small residual contribution.}
The analytic structure of eq.~\eqref{eq:F_II_IV_OPE} significantly differs from that of eq.~\eqref{eq:FI_OPE}. The logarithm develops a branch cut in the complex variable $s=p^2$ for $s>m_c^2$, while the pole singularity
associated with the denominator generates a cut in the variable
$s^\prime=(q-k)^2$, starting at
\begin{equation}
    \tilde s^\prime_{\rm III}(u, P^2) = \frac{- u P^2 + u \bar u m_\pi^2}{\bar u} \,,
\end{equation}
which depends on both $P^2$ and $u$. For large negative values of $P^2$, the non-vanishing contribution to the sum rule in eq.~\eqref{eq:SR} is
restricted to a small region around the endpoint $u=0$, with
\begin{equation}
u \leq -s_0^\pi/P^2\,, \qquad P^2\ll 0\,,
\end{equation} 
where corrections of order ${\cal O}\big((s_0^\pi/P^2)^2 \big)$ have been neglected.
In order to  be able to perform the $u$ integration analytically, 
we approximate those functions which are sufficiently smooth around $u=0$ with their endpoint limit and integrate only the remaining exponential dependence, 
still keeping $P^2$ large and negative.~\footnote{With this approximation, only a subset of the coefficients listed in appendix~\ref{sec:app2} contribute.} 
The result is an analytic function of $P^2$, which we can now analytically continue to the timelike point $P^2=m_D^2$, yielding
\begin{equation}
\begin{aligned}
&\int_0^{s_0^\pi} ds'\, e^{-s'/M'^2}\,
\operatorname{Im}_{s'}
F_{\mathrm{III}}^{Q_5}(s',s,m_D^2)\big|_{\mathrm{QL}} =
\pi f_\pi m_c
\ln\!\left(\frac{m_c^2-s-i \varepsilon}{m_c^2}\right)
\frac{M'^2}{m_D^2}
\left(e^{-s_0^\pi/M'^2}-1\right)
\\[2mm]
&\times
\sum_{\phi=\phi_{3\pi}^{p},\,\phi_{3\pi}^{\sigma}}
\left[
\phi(u)\,c_{\phi,1,\mathrm{III}}^{Q_5}(u,P^2,s)
+
\frac{d}{du}
\left(
\frac{
\phi(u)\,c_{\phi,2,\mathrm{III}}^{Q_5}(u,P^2,s)
}{
J_{\mathrm{III}}(u,P^2)
}
\right)
\right]_{\begin{subarray}{l}
u=0\\
P^2=m_D^2
\end{subarray}}
\left[
1+\mathcal{O}\!\left(\frac{s_0^\pi}{m_D^2}\right)
\right],
\end{aligned}
\label{eq:Q5_annihilation}
\end{equation}
where we have introduced the Jacobian
\begin{equation}
J_{\mathrm{III}}(u,P^2)
\equiv
\frac{\partial \tilde{s}'_{\mathrm{III}}(u,P^2)}
     {\partial u}
=
-\frac{P^2}{(1-u)^2}+m_\pi^2 \,.
\label{eq:annihilation-jacobian}
\end{equation}
The remaining integration over $s$ can then be performed analytically. 
An analogous expression holds for $D^0 \to K^+ K^-$, after the corresponding replacements. 
Note that in this case additional
corrections ${\cal O}(m_K^2/m_D^2)$ are also, for consistency, neglected.
The expression in eq.~\eqref{eq:Q5_annihilation} is consistent with the result obtained in~\cite{Khodjamirian:2005wn} for $B \to \pi\pi$. We stress however that, while in that case, the neglected corrections are only at the percent level, since
$s_0^\pi/m_B^2 \simeq 0.03$, for $D$ decays, the smaller charm-meson mass significantly increases the expansion parameters to 
\begin{equation}
s_0^\pi/m_D^2 \sim s_0^K/m_D^2 ={\cal O} (20 \mbox{-} 30\%)\,,
\end{equation}
so that the omitted terms are expected to be numerically more important. Consequently, the result in eq.~\eqref{eq:Q5_annihilation} should be viewed as a leading approximation with an expected uncertainty of at least 30\%.

Finally, we comment again on the difference that would arise if we had not introduced the auxiliary four-momentum $k$ in the correlation function. In this case, eq.~\eqref{eq:F_II_IV_OPE} would instead take the form:
\begin{align}
     \hat F_{{\rm II}}^{O}  \big(q^2,p^2 \big)_{\rm QL} = f_\pi m_c   \int_0^1   d u  \sum_\phi \phi(u)  \sum_{n = 1}^2  \frac{c_{\phi,n, {\rm II} }^{O} \big( u, p^2\big)}{\big( \hat{\tilde s}^\prime_{\rm II}   (u, p^2) - q^2 - i \varepsilon\big)^n} 
     \ln\left( \frac{m_c^2 - p^2 - i \varepsilon}{m_c^2}  \right)\,,
\end{align}
with
\begin{equation}
    \hat{\tilde s}^\prime_{\rm II}   (u, p^2) = 
  \frac{- \bar u p^2 + u \bar u m_\pi^2}{u}\,, 
\end{equation}
where for simplicity we have set $X =$ II. As before, the dependence on $p^2$ and $q^2$ is mixed, making the construction of a dispersion relation in $s=p^2$ non-trivial. However, unlike the tree-level topology, the function $\hat{\tilde s}^\prime_{\rm II}(u,p^2)$ appearing in the denominator generates an unphysical cut in the variable $s$, starting below the physical threshold $s=m_c^2$. This contribution is not parametrically suppressed and therefore spoils the derivation of the dispersion relations. 
The introduction of the auxiliary momentum $k$ avoids this problem by separating the two dispersion variables. 
We therefore conclude that in the approach used in~\cite{Lenz:2023rlq}, which did not include the four-momentum $k$, such contribution could not be consistently captured. Nevertheless, for current-current operators these annihilation effects first appear through subleading twist-four LCDAs and are found to give only a small contribution.
%%%%%%%%%%%%%%%%%%%%%%%%%%%%%%%%%%%%%%%%%%%%%%%%
%%%%%%%%%%%%%%%%%%%%%%%%%%%%%%%%%%%%%%%%%%%%%%%%
\subsection{Contributions from quark condensates}
\begin{figure}[t]
\centering

\includegraphics[width=0.33\textwidth]{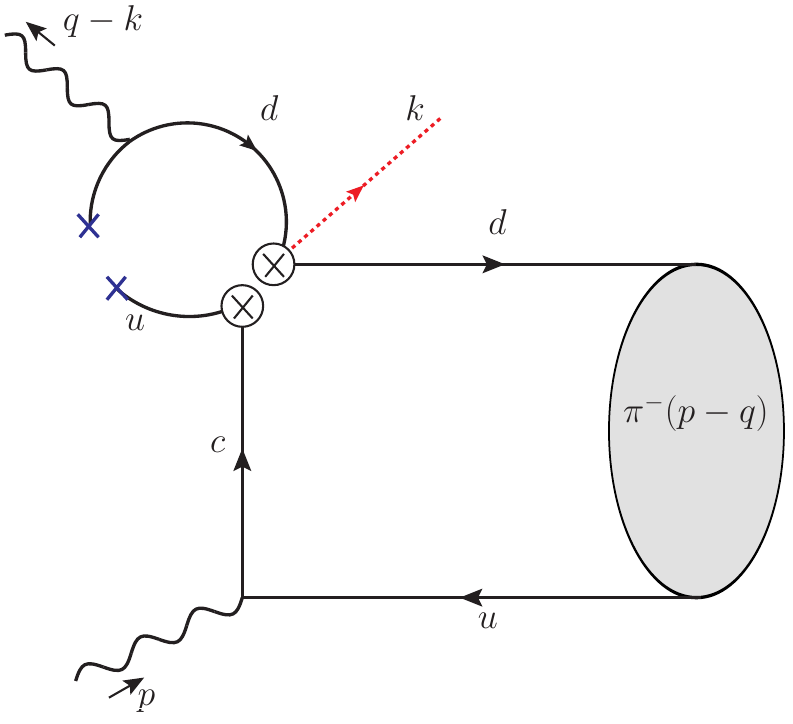}
\qquad
\includegraphics[width=0.33\textwidth]{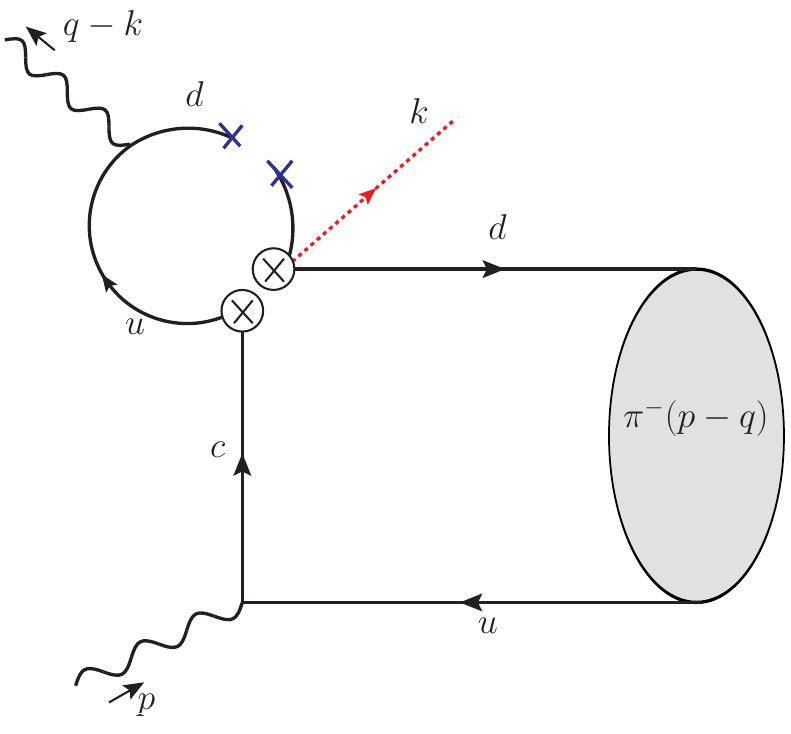}

\caption {Diagrammatic representation of the quark-condensate contribution to the correlation function in eq.~(\ref{eq:T1-QC}). The left (right) panel indicates the contribution from the up (down)
quark condensates.}
\label{fig:T1QC}

\end{figure}
The second class of contributions to the light-cone expansion of the correlation function in eq.~\eqref{eq:Correlator-1} is obtained by leaving, in addition to the external $\bar d u$ pair, one of the light-quark bilinears, $u\bar u$ or $d\bar d$, uncontracted in the time-ordered product. The resulting non-local matrix elements then involve four quark fields. For example, one such contribution takes the form
\begin{equation}
    \langle \pi^-(p-q)| \bar d^j_\alpha (0) [\ldots]_{\alpha \beta} \, u^j_\beta(y) \bar u^i_\delta(0) [\ldots]_{\delta \eta} u^i_\eta(x)|0\rangle\,,
\end{equation}
where $[\ldots]_{\alpha \beta}$ denotes a string of Dirac matrices with indices $\alpha, \beta$.

Following~\cite{Khodjamirian:2002pk}, see also e.g.~\cite{Braun:1999uj, Khodjamirian:1997tk, Rusov:2017chr}, we approximate these four-quark matrix elements by factorising them into a two-particle pion-to-vacuum matrix element and a vacuum-to-vacuum matrix element of the remaining $\bar u u $ pair. The latter is then expanded in terms of quark-condensates contributions.
Taking into account all possible light-quark field contractions, as well as colour and Dirac indices, two diagrams are relevant for our analysis, as shown in figure~\ref{fig:T1QC}.~\footnote{Additional contractions corresponding to annihilation-type topologies are in principle possible, but we find them to be numerically smaller. Here, we focus on the dominant contribution.}
The resulting contribution is non-vanishing only for the scalar penguin operator $Q_5$ due to its $(V-A)\otimes(V+A)$ Dirac structure. The corresponding diagrams for $Q_2^q$ and $Q_3$ vanish at leading order in $\alpha_s$ due to their Dirac structure in the limit of vanishing light-quark masses.

Using the condensate expansion given in appendix~\ref{sec:app1} and summing over all possible contractions, the quark-condensate contribution to the correlation function from $Q_5$ is
\begin{align}
F^{Q_5}_\mu (p, q, k)_{\rm QC} = - i  \frac{m_c}{4 N_c} &\int d^4 x \, e^{- i p \cdot x} \int d^4 y
\, e^{i (q-k) \cdot y} \, \Bigl\{
\langle \pi^- (p - q)|\biggl[\bar d(0) \Gamma_2^{Q_5} S_0^{(d)}(-y) 
\nonumber \\
& \times \gamma_\mu \gamma_5 \left(\langle \bar u u\rangle - \frac{i m_u}{4} \slashed{y} \langle \bar u u\rangle \right) \Gamma_1^{Q_5} S_0^{(c)}(-x) \gamma_5 u(x)\biggr]|0\rangle 
\nonumber\\
& + \langle \pi^- (p - q)|\biggl[\bar d(0) \Gamma_2^{Q_5} \left(\langle \bar d d\rangle + \frac{i m_d}{4} \slashed{y} \langle \bar d d\rangle \right)
\nonumber\\
& \times \gamma_\mu \gamma_5 S_0^{(u)}(y) \Gamma_1^{Q_5} S_0^{(c)}(-x) \gamma_5 u(x)\biggr] |0\rangle  \Bigr\}\,,
\label{eq:T1-QC}
\end{align}
where the two terms in the sum arise from including the quark-condensate contributions to the up- and down-quark bilinears, respectively, with $\langle \bar u u\rangle$ and $\langle \bar d d\rangle$ denoting the corresponding local condensates. 
Note that in eq.~\eqref{eq:T1-QC}, we retain for completeness the dependence on the light-quark masses, although numerically we set them to zero. 

The calculation of the condensate contribution in eq.~\eqref{eq:T1-QC} involves only integrals of single propagators and is therefore straightforward.
After inserting the light-cone expansion of the pion two-particle matrix element given in eq.~\eqref{eq:2pME}, the integrations over $x$ and $y$ are performed directly in coordinate space, using the representation in terms of Bessel functions for the charm-quark propagator, 
see e.g.~\cite{Piscopo:2023opf}, and its massless limit for the remaining light-quark propagators. 
The resulting invariant amplitude reads
\begin{equation}
F^{Q_5}\big( (q-k)^2, p^2, P^2 \big)_{\rm QC} = \mu_\pi  m_c f_\pi^3  \int_0^1 du \sum_{\phi} \phi(u) \sum_{n = 1}^3 \frac{c_{\phi, n, \rm{QC}}^{Q_5}(u)}{\big((q-k)^2 + i \varepsilon \big) \big(\tilde s_{\rm I}(u) - p^2 - i \varepsilon \big)^n}\,,
\label{eq:FqQCI-OPE}
\end{equation}
where the first sum runs over the pion LCDAs of twists two, three, four, $\tilde s_{\rm I}(u)$ is given in eq.~\eqref{eq:stilde_I} and the coefficients $c_{\phi, n, \rm{QC}}^{Q_5}$ are collected in appendix~\ref{sec:app2}. The overall proportionality to the chirally enhanced parameter $\mu_\pi$ follows from using the relation in eq.~\eqref{eq:QC-qq-relation}. 

This invariant amplitude exhibits pole singularities in both dispersion variables, $s=p^2$ and $s'=(q-k)^2$ and its imaginary part is obtained using the results listed in appendix~\ref{sec:app1}. Had we not introduced the auxiliary four-momentum $k$ in the correlation function, the corresponding result would be obtained from eq.~\eqref{eq:FqQCI-OPE} by replacing $(q-k)^2\to q^2$ and $\tilde s_{\rm I}(u)\to\hat{\tilde s}_{\rm I}(u,q^2)$, where the latter is defined in eq.~\eqref{eq:stilde_nok}. As in the case of the tree-level topology discussed above, the additional cut induced by $\hat{\tilde s}_{\rm I}(u,q^2)$ would lie entirely above the continuum threshold $s_0^\pi$ and therefore would not affect the sum rule. Furthermore, since eq.~\eqref{eq:FqQCI-OPE} is independent of $P^2$, no analytic continuation is required, and both correlation functions produce identical numerical results.
%%%%%%%%%%%%%%%%%%%%%%%%%%%%%%%%%%%%%%%%%%%%%%%%%
%%%%%%%%%%%%%%%%%%%%%%%%%%%%%%%%%%%%%%%%%%%%%%%%%
%%%%%%%%%%%%%%%%%%%%%%%%%%%%%%%%%%%%%%%%%%%%%%%%%
\section{Numerical analysis and results}
\label{sec:Results}
In this section, we discuss the numerical analysis and present our results for the hadronic matrix elements of the four-quark operators in eqs.~\eqref{eq:Q1} - \eqref{eq:Q5} computed within the LCSR framework at leading order in the strong coupling $\alpha_s$. 
 Using our results for the current-current operators, we update the LCSR predictions of the branching ratios of the singly-Cabibbo suppressed modes $D^0 \to \pi^+ \pi^-$ and $D^0 \to K^+ K^-$, comparing them with the values previously obtained in ~\cite{Lenz:2023rlq}. Moreover, combining our results for the matrix elements of the QCD penguin operators with the available LCSR determinations of the matrix elements of the penguin contractions of the current-current operators~\cite{Khodjamirian:2017zdu}, we discuss the impact of the penguin operators on the size of direct CP violation in charm and derive a new bound on its expected magnitude within the SM. 
\subsection{Discussion of the input used}
We start by discussing the input parameters used in our analysis, which are also conveniently summarised in table~\ref{tab:input}. 
Our choice of input closely follows the LCSR analysis of $D^0\to K^+K^-$ and $D^0\to\pi^+\pi^-$ presented in~\cite{Lenz:2023rlq}. Specifically, 
for the parametrisation of the LCDAs up to twist four, we use the results of~\cite{Duplancic:2008ix} for the pion case and those of~\cite{Khodjamirian:2009ys}, based on~\cite{Ball:2007zt}, for the kaon. 
These parametrisations are expressed as Gegenbauer expansions with coefficients $a_n^\pi$ and $a_n^K$. 
For the pion, the odd Gegenbauer moments vanish by isospin symmetry and we neglect moments with $n>4$, while the numerical values of the remaining coefficients are taken from~\cite{Cheng:2020vwr}. 
For the kaon, we retain the first two Gegenbauer moments and neglect $a_n^K$ with $n>2$. Their numerical values are taken from~\cite{Chetyrkin:2007vm,Ball:2007zt}. 
The chirally enhanced parameters $\mu_\pi$ and $\mu_K$, entering the twist-three LCDAs, are taken from~\cite{Cheng:2020vwr}, while the remaining twist-three and twist-four parameters are taken from~\cite{Ball:2007zt}. 
These scale-dependent non-perturbative input parameters are evolved from the scale at which they are determined, typically 1 GeV, using leading-order renormalisation-group equations, whose explicit expressions can be found, for example, in the appendix of~\cite{Khodjamirian:2009ys}. 

As for the remaining non-perturbative input, the meson decay constants are precisely known from lattice QCD determinations and we use the averages reported by FLAG~\cite{FLAG:2024oxs}. 
The running of the strong coupling and of the quark masses in the $\overline{\rm MS}$ scheme is implemented using the Mathematica package \texttt{RunDec}~\cite{Herren:2017osy} with two-loop accuracy. 
For the Wilson coefficients entering the branching-ratio and CP-asymmetry predictions, we consider both the LO and NLO values~\cite{Buchalla:1995vs} listed in table~\ref{tab:wc}. 
The renormalisation scale is varied in the range $1 \leq \mu \leq 2$ GeV, taking $\mu=1.5$ GeV as the central value. 
We use the standard parametrisation of the CKM matrix together with the input values from the global CKM fit provided by the CKMfitter collaboration~\cite{Charles:2004jd} (online update), which are consistent with those obtained by the UTfit collaboration~\cite{UTfit:2022hsi}.

For the sum-rule parameters, we adopt the same values as in~\cite{Lenz:2023rlq}, largely following~\cite{Khodjamirian:2017zdu}. In the pion and kaon channels we use the Borel parameters $M^{\prime2} \equiv M_\pi^2=(1.0\pm0.5)\,\mathrm{GeV}^2$ and $M^{\prime2} \equiv M_K^2=(1.0\pm0.5)\,\mathrm{GeV}^2$ together with the continuum thresholds $s_0^\pi=(0.7\pm0.1)\,\mathrm{GeV}^2$ and $s_0^K=(1.2\pm0.1)\,\mathrm{GeV}^2$~\cite{Braun:1999uj,Bijnens:2002mg}. For the $D$-meson channel we adopt $M^2\equiv M_D^2=(4.5\pm1.0)\,\mathrm{GeV}^2$~\cite{Khodjamirian:2017zdu} and $s_0^D=(8.2^{+1.4}_{-0.6})\,\mathrm{GeV}^2$~\cite{Lenz:2023rlq}. The dependence of the results on these sum-rule parameters is discussed below.
\renewcommand{\arraystretch}{1.2}
\begin{table}[t]
\centering
\setlength{\tabcolsep}{4pt}
\begin{tabular}{ccl|ccl}
\toprule
\multicolumn{6}{c}{Pion and kaon twist-two and twist-three LCDAs parameters} \\
\midrule

$\mu_\pi$ & $(2.50 \pm 0.30)\,\GeV$ & \cite{Khodjamirian:2017fxg} &
$\mu_K$ & $(2.49 \pm 0.26)\,\GeV$ & \cite{Khodjamirian:2017fxg} \\

$a_2^\pi$ & $0.275 \pm 0.055$ & \cite{Cheng:2020vwr} &
$a_1^K$ & $0.10 \pm 0.04$ & \cite{Chetyrkin:2007vm} \\

$a_4^\pi$ & $0.185 \pm 0.065$ & \cite{Cheng:2020vwr} &
$a_2^K$ & $0.25 \pm 0.15$ & \cite{Ball:2007zt} \\

$f_3^\pi$ & $(0.0045 \pm 0.0015)\,\GeV^2$ & \cite{Ball:2007zt} &
$f_3^K$ & $(0.0045 \pm 0.0020)\,\GeV^2$ & \cite{Ball:2007zt} \\

$\omega_3^\pi$ & $-1.5 \pm 0.7$ & \cite{Ball:2007zt} &
$\omega_3^K$ & $-1.2 \pm 0.7$ & \cite{Ball:2007zt} \\

$\lambda_3^\pi$ & $0$ & --- &
$\lambda_3^K$ & $1.6 \pm 0.4$ & \cite{Ball:2007zt} \\

\midrule
\multicolumn{6}{c}{Pion and kaon twist-four LCDAs parameters} \\
\midrule

$\delta_\pi^2$ & $(0.18 \pm 0.06)\,\GeV^2$ & \cite{Ball:2007zt} &
$\delta_K^2$ & $(0.20 \pm 0.06)\,\GeV^2$ & \cite{Ball:2007zt} \\

$\omega_4^\pi$ & $0.20 \pm 0.10$ & \cite{Ball:2007zt} &
$\omega_4^K$ & $0.20 \pm 0.10$ & \cite{Ball:2007zt} \\

$\kappa_{4\pi}$ & $0$ & --- &
$\kappa_{4K}$ & $-0.12 \pm 0.01$ & \cite{Ball:2007zt} \\

\midrule
\multicolumn{6}{c}{Sum-rule parameters} \\
\midrule

$s_0^\pi$ & $(0.7 \pm 0.1)\,\GeV^2$ & \cite{Khodjamirian:2017zdu} &
$s_0^K$ & $(1.2 \pm 0.1)\,\GeV^2$ & \cite{Khodjamirian:2017zdu} \\

$M_\pi^2$ & $(1.0 \pm 0.5)\,\GeV^2$ & \cite{Khodjamirian:2006st} &
$M_K^2$ & $(1.0 \pm 0.5)\,\GeV^2$ & \cite{Khodjamirian:2006st} \\

$s_0^D$ & $8.2^{+1.4}_{-0.6}\,\GeV^2$ & \cite{Lenz:2023rlq} &
$M_D^2$ & $(4.5 \pm 1.0)\,\GeV^2$ & \cite{Khodjamirian:2009ys} \\

\midrule
\multicolumn{6}{c}{CKM parameters} \\
\midrule

$|V_{us}|$ & $0.22500^{+0.00024}_{-0.00021}$ & \cite{Charles:2004jd} &
$\dfrac{|V_{ub}|}{|V_{cb}|}$ & $0.08848^{+0.00224}_{-0.00219}$ &
\cite{Charles:2004jd} \\

$|V_{cb}|$ & $0.04145^{+0.00035}_{-0.00061}$ & \cite{Charles:2004jd} &
$\delta$ & $\left(65.5^{+1.3}_{-1.2}\right)^\circ$ &
\cite{Charles:2004jd} \\

\midrule
\multicolumn{6}{c}{Other parameters} \\
\midrule

$m_{\pi^\pm}$ & $0$ & --- &
$m_{K^\pm}$ & $0.493677\,\GeV$ & \cite{ParticleDataGroup:2026aaa} \\

$f_\pi$ & $(0.1302 \pm 0.0008)\,\GeV$ & \cite{FLAG:2024oxs} &
$f_K$ & $(0.1557 \pm 0.0003)\,\GeV$ & \cite{FLAG:2024oxs} \\

$\overline{m}_c$ & $(1.27 \pm 0.02)\,\GeV$ & \cite{ParticleDataGroup:2026aaa} &
$m_{u,d,s}$ & 0 &
--- \\

$\alpha_s(m_Z)$ & $0.1179 \pm 0.0009$ & \cite{ParticleDataGroup:2026aaa} &
$m_{D^0}$ & $1.86484\,\GeV$ & \cite{ParticleDataGroup:2026aaa} \\

$f_D$ & $(0.2120 \pm 0.0007)\,\GeV$ & \cite{FLAG:2024oxs} &
$\tau(D^0)$ & $(0.4103 \pm 0.0010)\,\mathrm{ps}$ &
\cite{ParticleDataGroup:2026aaa} \\
\bottomrule
\end{tabular}
\caption{
Inputs used in the numerical analysis. The values of all non-perturbative parameters entering the LCDAs, as well as of the quark masses correspond to the renormalisation scale $\mu=1$ GeV, apart from the chirally-enhanced parameters $\mu_\pi$ and $\mu_K$, which correspond to the scale $\mu=2$ GeV. We use $\gamma=\delta$, neglecting the numerically tiny difference between the two phases~\cite{ParticleDataGroup:2026aaa}.
}
\label{tab:input}
\end{table}
%%%%%%%%%%%%%%%%%%%%%%%%%%%%%%%%%%%%%%%%%%%%%%%%%%%%%%%%%%%%%%%%%
%%%%%%%%%%%%%%%%%%%%%%%%%%%%%%%%%%%%%%%%%%%%%%%%%%%%%%%%%%%%%%%%%
%%%%%%%%%%%%%%%%%%%%%%%%%%%%%%%%%%%%%%%%%%%%%%%%%%%%%%%%%%%%%%%%%
\subsection{Results for the matrix elements and branching ratios}
We now present our numerical results. Starting from the LCSR determinations of the matrix elements of the operators $Q_2^q$, $Q_3$, and $Q_5$, outlined in section~\ref{sec:OPE}, the remaining matrix elements can be related using the identity in eq.~\eqref{eq:Fierz_id}, see appendix~\ref{sec:app1p}. 
The total contribution from the current-current operators to the decay amplitudes ${\cal A}_{\pi\pi}$ and ${\cal A}_{KK}$ in eqs.~\eqref{eq:Apipi}, \eqref{eq:Akk} reduces to
\begin{equation}
{\cal A}_{\pi\pi} = -\frac{G_F}{\sqrt{2}}   
\left[ \left( C_1 + \frac{C_2}{N_c}\right) \langle Q_1^d \rangle_{\pi\pi}^{\rm P_{\rm I}} + \left( C_2 + \frac{C_1}{N_c}\right) N_c 
\langle Q_1^d \rangle_{\pi\pi}^{\rm P_{\rm II}} \right]\,, 
\label{eq:iApipi}
\end{equation}
and
\begin{equation}
{\cal A}_{KK} =- \frac{G_F}{\sqrt{2}}   
\left[ \left( C_1 + \frac{C_2}{N_c}\right) \langle Q_1^s \rangle_{KK}^{\rm P_{\rm I}}  + \left( C_2 + \frac{C_1}{N_c}\right)  N_c \langle Q_1^s \rangle_{KK}^{\rm P_{\rm II}}  \right] \,, 
\label{eq:iAKK}
\end{equation}
where $\langle\cdots\rangle_{\pi\pi}\equiv\langle\pi^+\pi^-|\cdots|D^0\rangle$, with the superscript indicating the contribution from the corresponding topology, and analogously for the $K^+ K^-$ final state.
As for the CKM subleading amplitudes ${\cal P}_{\pi\pi}$ and ${\cal P}_{KK}$ in eq.~\eqref{eq:P}, they receive contributions from the tree-level matrix elements of the QCD penguin operators, determined in this work, as well as from the penguin contractions of the current-current operators obtained in~\cite{Khodjamirian:2017zdu}. We write
\begin{align}
{\cal P}_{\pi\pi} = -\frac{G_F}{\sqrt{2}} \Bigg[ 
{\cal P}_{\pi\pi}^s + 
&\left( C_4 +  \frac{C_3}{N_c}\right) \langle Q_4 \rangle_{\pi\pi}^{\rm P_{\rm I} + P_{\rm III}} +
\left( C_6 + \frac{C_5}{N_c}\right)
\langle Q_6 \rangle_{\pi\pi}^{\rm P_{\rm III} + QC} 
\nonumber \\[2mm]
& + \left( C_3 +  \frac{C_4}{N_c}\right) N_c \langle Q_4 \rangle_{\pi\pi}^{\rm P_{\rm II} + P_{\rm IV}} +
\left( C_5 + \frac{C_6}{N_c}\right)
N_c  \langle Q_6 \rangle_{\pi\pi}^{\rm P_{\rm II} + P_{\rm IV}}
\Bigg]\,,
\label{eq:iPpipi}
\end{align}
and
\begin{align}
{\cal P}_{KK} = - \frac{G_F}{\sqrt{2}} \Bigg[ 
{\cal P}_{KK}^d 
+
& \left( C_4 + \frac{C_3}{N_c}\right) \langle Q_4 \rangle_{KK}^{\rm P_{\rm I} + P_{\rm III}}
+
\left( C_6 + \frac{C_5}{N_c}\right) \langle Q_6 \rangle_{KK}^{\rm P_{\rm III} + QC} 
\nonumber 
\\[2mm]
& + \left( C_3 +  \frac{C_4}{N_c}\right) N_c \langle Q_4 \rangle_{KK}^{\rm P_{\rm II} + P_{\rm IV}} +
\left( C_5 + \frac{C_6}{N_c}\right)
N_c\langle Q_6 \rangle_{KK}^{\rm P_{\rm II} + P_{\rm IV}}
\Bigg]\,.
\label{eq:iPKK}
\end{align}
Here, the notation for the penguin contraction of the current-current operators follows that in~\cite{Khodjamirian:2017zdu}, i.e.
\begin{equation}
    {\cal P}_{\pi\pi}^s \equiv 2 C_1 \langle \pi^+ \pi^-| \tilde Q_2^s| D^0\rangle\,, \qquad
     {\cal P}_{KK}^d \equiv 2 C_1 \langle K^+ K^-| \tilde Q_2^d| D^0\rangle\,,
\end{equation}
where 
\begin{equation}
    \tilde Q_2^q =  \left(\bar q^i t^a_{ij} q^j \right)_{V-A}  \left(\bar u^k t^a_{km}  c^m \right)_{V-A}\,, \qquad q = d,s\,,
    \label{eq:octet_op}
\end{equation}
is the colour-octet operator.

Our LCSR predictions for the hadronic matrix elements of the current-current and QCD penguin operators at leading order in $\alpha_s$ are summarised in table~\ref{tab:matrixelements}. For each operator, we display separately the central values of the quark-loop and quark-condensate contributions, together with the total result and its uncertainties.
The first uncertainty combines the variations of the input parameters and the renormalisation scale. Specifically, each input parameter is varied individually around its central value while keeping the remaining inputs fixed, and the resulting shifts are added in quadrature. The scale uncertainty is obtained by varying $\mu$ around its central value and is then added in quadrature to the parametric error. The second uncertainty is a conservative estimate of effects beyond the accuracy of the calculation, including missing higher-order perturbative corrections, higher-twist effects, and other neglected power corrections; we assign it as $40 \%$ of the central value. 

First, we discuss the main sources of theoretical uncertainties, starting with the parametric ones. For the $\pi^+ \pi^-$ final state, the current-current operator $Q_1^d$ and the QCD
penguin operator $Q_4$ exhibit the largest sensitivity to the
Borel parameter $M_\pi^2$, which changes the corresponding
matrix elements by approximately $11$--$25\%$ within the adopted
Borel window.
The continuum threshold $s_0^\pi$ and the chiral parameter
$\mu_\pi$ induce somewhat smaller variations, typically at the
$8$--$18\%$ level. 
In contrast, for the penguin operator $Q_6$, the
dominant uncertainties arise from the $D$-meson continuum threshold
$s_0^D$, the pion Borel parameter $M_\pi^2$, and the chiral parameter
$\mu_\pi$, each modifying the matrix element by roughly
$8$--$17\%$. The dependence on the remaining sum-rule parameters,
including $M_D^2$, as well as on the higher-twist LCDA input, is
generally below the $7\%$ level.
\begin{table}[!t]
\centering
\renewcommand{\arraystretch}{1.8}
\begin{tabular}{c|ccccc|c}
\toprule
& $\mathrm{P}_{\rm I}$
& $\mathrm{P}_{\rm II}$
& $\mathrm{P}_{\rm III}$
& $\mathrm{P}_{\rm IV}$
& QC
& Total \\
\hline
\hline
$-i\langle Q_1^d\rangle_{\pi\pi}$
& $0.221$
& $0.007$
& --
& --
& $0$
& $0.228^{+0.042}_{-0.072}\pm0.091$
\\
\hline
$-i\langle Q_4\rangle_{\pi\pi}$
& $0.221$
& $0.007$
& $-0.022$
& $-0.007$
& $0$
& $0.199^{+0.041}_{-0.067}\pm0.080$
\\
\hline
$-i\langle Q_6\rangle_{\pi\pi}$
& $0$
& $0.007$
& $1.13$
& $-0.007$
& $0.955$
& $2.09^{+0.55}_{-0.62}\pm0.83$
\\
\hline
\hline
$-i\langle Q_1^s\rangle_{KK}$
& $0.342$
& $0.032$
& --
& --
& $0$
& $0.374^{+0.052}_{-0.087}\pm0.150$
\\
\hline
$-i\langle Q_4\rangle_{KK}$
& $0.342$
& $0.032$
& $-0.046$
& $-0.015$
& $0$
& $0.313^{+0.051}_{-0.082}\pm0.125$
\\
\hline
$-i\langle Q_6\rangle_{KK}$
& $0$
& $0.032$
& $1.67$
& $-0.015$
& $1.79$
& $3.48^{+0.86}_{-0.96}\pm1.39$
\\
\bottomrule
\end{tabular}
\caption{
LCSR predictions in units of $\mathrm{GeV}^3$ for the hadronic
matrix elements of the current-current and QCD penguin operators
$Q_1^q$, $Q_4$, and $Q_6$ for the $D^0\to\pi^+\pi^-$ and
$D^0\to K^+K^-$ decay modes, obtained at leading order in $\alpha_s$.
For each operator, we show separately the contributions from the
individual quark-loop topologies I--IV and from quark condensates
(QC), together with their sum.
The central values correspond to input parameters evaluated at
the renormalisation scale $\mu=1.5\,\mathrm{GeV}$.
In the total results, the first uncertainty is obtained by adding
in quadrature the variations of all input parameters and of the
renormalisation scale, while the second uncertainty estimates
missing higher-order QCD corrections and is taken to be
$\pm40\%$ of the central value.
}
\label{tab:matrixelements}
\end{table}

A similar pattern is observed for the kaon channel. For the current-current
operator $Q_1^s$, the dominant uncertainty originates from the
light-meson Borel parameter $M_K^2$, which changes the matrix element by
approximately $16$--$24\%$, while the dependence on the chiral parameter
$\mu_K$ is at the level of about $7\%$. For the QCD penguin operator $Q_4$, the largest uncertainties arise from
the kaon-channel Borel parameter $M_K^2$, which changes the matrix
element by approximately $8$--$16\%$, and from the first Gegenbauer
moment of the kaon LCDA, $a_1^K$, which induces an uncertainty of about
$\pm10\%$.
Finally, for the penguin operator $Q_6$, the
dominant parametric uncertainties originate from the chiral parameter
$\mu_K$ and the $D$-meson continuum threshold $s_0^D$, each inducing
variations of approximately $8$--$16\%$, whereas the dependence on the
remaining sum-rule parameters is comparatively mild. 
In addition, the uncertainty due to renormalisation-scale variation is sizeable for all matrix elements. It amounts to
approximately $7$--$15\%$ for the current-current operator $Q_1^q$,
$4$--$18\%$ for the QCD penguin operator $Q_4$, and
$10$--$21\%$ for the penguin operator $Q_6$.

We now turn to discussing our findings, starting with the matrix elements of the colour-singlet current-current operators $Q_1^d$ and $Q_1^s$. In this case, the annihilation contribution is numerically suppressed, since, as discussed above, it first arises through subleading twist-four LCDAs. 
Comparing the values in table~\ref{tab:matrixelements} with those obtained in~\cite{Lenz:2023rlq}, 
we find very good agreement, although our results are found to be systematically smaller, respectively by $\sim 5\%$ and $\sim 10\%$ at the central value level. This close agreement shows that the two different correlation functions used in the present and the previous work~\cite{Lenz:2023rlq} lead only to a mild difference for the matrix elements of the current-current operators at leading order in $\alpha_s$, which is consistent with the dominance of the tree-level topology at this order. This comparison provides both an independent determination of these matrix elements and an important consistency check of the LCSR framework.
\begin{table}[!t]
\centering
\renewcommand{\arraystretch}{1.8}
\begin{tabular}{c|cc}
\toprule
& \mbox{This work} & \mbox{HFLAV~\cite{HFLAV:2024ctg}}
\\
\hline
\hline
$10^3 \times {\cal B}(D^0\to \pi^+\pi^-)$
& $1.22^{+1.26}_{-1.04}$
& $1.490 \pm 0.027$ \\
$10^3 \times {\cal B}(D^0\to K^+K^-)$
& $2.44^{+2.45}_{-1.91}$
& $4.113 \pm 0.051$ \\
\bottomrule
\end{tabular}
\caption{Comparison of our LCSR predictions for the branching ratios of
$D^0 \to \pi^+\pi^-$ and $D^0 \to K^+K^-$ with the corresponding
experimental averages.}
\label{tab:BR}
\end{table}

Using the expressions in eq.~\eqref{eq:Br} together with LO values for the Wilson coefficients, we obtain the results for the branching ratios shown in table~\ref{tab:BR}.~\footnote{For the branching fractions, the $40\%$ uncertainty is propagated through the squared decay amplitudes, corresponding to multiplicative factors of $(1+0.4)^2=1.96$ and $(1-0.4)^2=0.36$ for the upper and lower variations, respectively. As a result, the branching-ratio uncertainties exhibit a larger upward than downward shift, despite the corresponding matrix-element uncertainties showing an asymmetry in the opposite direction.} Our results are in agreement with the experimental averages~\cite{HFLAV:2024ctg} within uncertainties and are also consistent with the previous LCSR results~\cite{Lenz:2023rlq}. 
Compared with the latter analysis, however, our central values are smaller, as a consequence of the smaller central values of the current-current matrix elements discussed above, and of the fact that the contribution from topology II interferes destructively with that from topology I due to the combination of Wilson coefficients, cf.~eqs.~\eqref{eq:iApipi}, ~\eqref{eq:iAKK}. 
Our uncertainties are of comparable size and amount to about $100\%$.
We note that, had we used NLO values for the Wilson coefficients, the results in table~\ref{tab:BR} would change only mildly, by approximately $-3\%$ for the $\pi^+\pi^-$ final state and by less than $1\%$ for $K^+K^-$. This follows from the fact that the combination of Wilson coefficients in eqs.~\eqref{eq:iApipi}, \eqref{eq:iAKK} is rather stable upon including QCD corrections. 
Moreover, for the ratio of branching ratios and the size of $U$-spin breaking, we find
\begin{equation}
\left.
\frac{{\cal B}(D^0\to K^+K^-)}
     {{\cal B}(D^0\to \pi^+\pi^-)}
\right|_{\rm LCSR}
=
2.00^{+0.68}_{-0.60}\,,
\qquad
\left.
\frac{{\cal B}(D^0\to K^+K^-)}
     {{\cal B}(D^0\to \pi^+\pi^-)}
\right|_{\rm exp}
= 2.760 \pm 0.060\,,
\end{equation}
where we also compare with the current experimental determination. The latter is obtained using the individual averages of the branching ratios from~\cite{HFLAV:2024ctg} combined in the ratio without taking possible correlations into account. Our result is consistent with the experimental determination within uncertainties, which are reduced by approximately a factor of two when considering the ratio. The amount of $U$-spin breaking is smaller than that obtained in the previous LCSR analysis~\cite{Piscopo:2024wpd}. 
This difference can be attributed mainly to the fact that we do not include $m_s$ corrections in our analysis.

Moving to the QCD penguin operators, from table~\ref{tab:matrixelements} we see that the matrix element of $Q_4$ is of the same order of magnitude as that of the colour-singlet current-current operator $Q_1^q$, whereas the matrix element of $Q_6$ is enhanced by approximately a factor of ten. 
This enhancement can be traced to the $(V-A) \otimes (V+A) $ Dirac structure of $Q_6$, which allows the operator to couple directly to the chirally enhanced pseudoscalar light-meson LCDAs. As a consequence, on the one hand the quark-condensate contribution is non-vanishing and sizeable. We stress that although these condensate terms formally correspond to subleading twist-five and -six contributions when combined with twist-two and twist-three LCDAs, respectively, see for details~\cite{Khodjamirian:2002pk, Rusov:2017chr}, they arise already at tree level, whereas the remaining contributions to the sum rule originate from one-loop diagrams. As a result, these condensate terms are enhanced by a relative factor of $16\pi^2$, in particular compared to the current-current operator contributions. 
On the other hand, the annihilation topology III already receives a contribution from twist-three LCDAs, yielding a sizeable effect of comparable magnitude to the quark-condensate term.

At the same time, the quark-condensate contribution vanishes for $Q_4$, at the order considered here, while the topology III is numerically suppressed because it receives contributions only from twist-four LCDAs. For all penguin operators, the contributions from topologies II and IV cancel identically for $D^0 \to \pi^+ \pi^-$, while their sum is suppressed, although non-vanishing, for $D^0 \to K^+ K^-$, as previously noted.

Our findings are consistent with the picture obtained in the naive factorisation approximation and discussed in section~\ref{sec:Heff}. Furthermore, they confirm the significant effect arising from the annihilation contribution of $Q_6$, observed in the LCSR analysis of $B \to \pi \pi$~\cite{Khodjamirian:2005wn}.

Despite the sizeable matrix element of $Q_6$, the QCD penguin operators have a negligible impact on the branching ratios. Their contributions are suppressed by the small Wilson coefficients, in particular through the specific combinations entering eqs.~\eqref{eq:iPpipi} and~\eqref{eq:iPKK}, and, most importantly, by the small CKM factor $\lambda_b$. This enhancement is nevertheless phenomenologically important for direct CP violation, as we discuss below.
%%%%%%%%%%%%%%%%%%%%%%%%%%%%%%%%%%%%%%%%%%%%%%%%%%%%%%%%%%%%%%%%%
%%%%%%%%%%%%%%%%%%%%%%%%%%%%%%%%%%%%%%%%%%%%%%%%%%%%%%%%%%%%%%%%%
%%%%%%%%%%%%%%%%%%%%%%%%%%%%%%%%%%%%%%%%%%%%%%%%%%%%%%%%%%%%%%%%%
\subsection{Results for the direct CP asymmetries}
In this section, we discuss the implications of our results for the matrix elements of the QCD penguin operators in table~\ref{tab:matrixelements} for the expected size of direct CP violation in $D^0\to\pi^+\pi^-$ and $D^0\to K^+K^-$. To this end, we use the expression given in
eq.~\eqref{eq:Delta_acp}. 

Within our approach, the only strong phases originate from the penguin contractions of the current-current operators, i.e. from ${\cal P}_{\pi\pi}^s$ and ${\cal P}_{KK}^d$ in eqs.~\eqref{eq:iPpipi}, \eqref{eq:iPKK}, which have been determined within the same LCSR framework in~\cite{Khodjamirian:2017zdu}. 
Our calculation of the matrix elements of the current-current and QCD penguin operators does not generate additional strong phases at the order ${\cal O}(\alpha_s^0)$ considered here. A determination of such phases would require a substantially more involved analysis and lies beyond the scope of this work. Nevertheless, it is instructive to assess the potential impact of the QCD penguin matrix elements on direct CP violation by assuming maximal relative strong phases, $\sin\phi_{\pi\pi}=\sin\phi_{KK}=1$. This assumption allows us to estimate the largest possible values of the asymmetries within the ranges of the hadronic uncertainties, following the approach adopted in~\cite{Khodjamirian:2017zdu,Lenz:2023rlq}.
\begin{table}[!t]
\centering
\renewcommand{\arraystretch}{1.35}
\setlength{\tabcolsep}{8pt}
\begin{tabular}{ccccc}
\toprule
& \multicolumn{2}{c}{Only $Q_1$, $Q_2$}
& \multicolumn{2}{c}{Full result} \\
\cmidrule(lr){2-3}\cmidrule(lr){4-5}
& LO-WC & NLO-WC & LO-WC & NLO-WC \\
\midrule
$\displaystyle
\left|\frac{\mathcal P_{\pi\pi}}{\mathcal A_{\pi\pi}}\right|
$
& $0.099^{+0.057}_{-0.043}$
& $0.096^{+0.054}_{-0.042}$
& $0.217^{+0.198}_{-0.119}$
& $0.348^{+0.313}_{-0.185}$
\\[4mm]
$\displaystyle
\left|\frac{\mathcal P_{KK}}{\mathcal A_{KK}}\right|
$
& $0.101^{+0.057}_{-0.043}$
& $0.096^{+0.052}_{-0.041}$
& $0.221^{+0.218}_{-0.125}$
& $0.358^{+0.335}_{-0.193}$
\\
\bottomrule
\end{tabular}
\caption{
Comparison of the absolute values of the ``penguin-to-tree'' amplitude
ratios obtained using the LCSR determination of the hadronic matrix
elements.
The second and third columns show the results obtained using only the
penguin contractions of the current-current operators
from~\cite{Khodjamirian:2017zdu}.
The fourth and fifth columns show the full results, including both the
penguin contractions of the current-current operators and the
contributions from the QCD penguin operators obtained in this work.
In each case, we compare the results obtained using LO and NLO Wilson
coefficients (WC).
The quoted uncertainties include an additional $40\%$ relative
uncertainty on each ratio, combined in quadrature with the other
uncertainties.
For simplicity, the superscript $d$ or $s$ on the current-current
operators is omitted.
}
\label{tab:phenomenology}
\end{table}

Combining our results for the matrix elements of the current-current and QCD penguin operators given in table~\ref{tab:matrixelements}, together with the matrix elements of the colour-octet operators $\tilde Q_2^d$ and $\tilde Q_2^s$ from~\cite{Khodjamirian:2017zdu}, we find for the magnitudes of the ratios ${\cal P}_{\pi\pi}/{\cal A}_{\pi\pi}$ and ${\cal P}_{KK}/{\cal A}_{KK}$ the values shown in table~\ref{tab:phenomenology}. 
A few comments are in order. 

First, we observe that the inclusion of the QCD penguin operators leads to a significant enhancement of these ratios. This is illustrated in table~\ref{tab:phenomenology} where we show the corresponding values obtained neglecting the contribution of the QCD penguins, as in~\cite{Khodjamirian:2017zdu, Lenz:2023rlq}, with the full result which includes both penguin contractions of the current-current operators and penguin operators matrix elements. 
Second, the results exhibit a significant sensitivity to the perturbative accuracy of the Wilson coefficients used. The amplitudes ${\cal A}_{\pi\pi,KK}$ and ${\cal P}_{\pi\pi,KK}^{s,d}$ are relatively stable when going from LO to NLO Wilson coefficients, owing to the stability of the combinations $C_1+ C_2/N_{c}$ and $C_1$. In contrast, the contribution from the QCD penguin operators is much more sensitive to this choice, since the corresponding combinations of Wilson coefficients entering eqs.~\eqref{eq:iPpipi}, \eqref{eq:iPKK} change substantially between LO and NLO. 
As illustrated in table~\ref{tab:phenomenology}, the inclusion of the QCD penguin operators enhances the ratios by approximately a factor of two when LO Wilson coefficients are used, and by approximately a factor of four when NLO Wilson coefficients are employed.

Since the contribution of the QCD penguin operators to the effective Hamiltonian first arises at order $\alpha_s$, it is formally consistent to treat it as a subleading contribution and use LO Wilson coefficients in a leading-order analysis. Nevertheless, varying the Wilson coefficients from LO to NLO while keeping the hadronic matrix elements fixed provides a useful estimate of the sensitivity to higher-order perturbative effects and, in particular, gives an indication of the potential size of corrections beyond the accuracy of our calculation.

We also note that, when only the penguin contractions of the current-current operators are included, our results for the ratios are in good agreement with those found in~\cite{Lenz:2023rlq}. Our central values are, however, about $10$--$40\%$ larger and exhibit less $U$-spin breaking. This difference can be attributed to two effects: first, our central values for the matrix elements of the current-current operators are smaller, as discussed above; second, we do not include explicit $m_s$ corrections, leading to more $U$-spin symmetric ratios between the $\pi^+ \pi^-$ and $K^+K^-$ final states.

As for the uncertainties of our results, these are obtained by varying all input parameters within their respective uncertainty ranges and assigning an additional $40\%$ uncertainty, analogously to the procedure used for the branching ratios. For the full result, the resulting uncertainties amount to about $50\%$ when using LO Wilson coefficients and up to about $100\%$ when using NLO Wilson coefficients. This further illustrates the need for a consistent treatment of higher-order corrections and of the renormalisation of the corresponding hadronic matrix elements.

Finally, combining our results for the magnitudes of ${\cal P}_{\pi\pi}/{\cal A}_{\pi\pi}$ and ${\cal P}_{KK}/{\cal A}_{KK}$, and assuming maximal relative strong phases, we obtain the following maximal values for the possible size of $\Delta a_{\rm CP}^{\rm dir}$, obtained within the quoted uncertainties of the penguin-to-tree ratios:
\begin{equation}
|\Delta a_{\rm CP}^{\rm dir}|_{\rm LCSR} \lesssim
\begin{cases}
11.1 \times 10^{-4} & \text{(LO-WC)},\\[2mm]
17.7 \times 10^{-4} & \text{(NLO-WC)},
\end{cases}
\qquad
|\Delta a_{\rm CP}^{\rm dir}|_{\rm exp}
= (15.7\pm2.9) \times 10^{-4},
\end{equation}
where the two results correspond to the use of LO and NLO Wilson coefficients, respectively, and the experimental value is taken from~\cite{LHCb:2019hro}. 
The quoted LCSR values should be interpreted as the largest values reached when the uncertainties of the penguin-to-tree ratios are taken into account.
For comparison, neglecting the contribution of the QCD penguin operators gives bounds of $\lesssim 4.1 \times 10^{-4}$ and $\lesssim 3.9 \times 10^{-4}$ using LO and NLO Wilson coefficients, respectively. These bounds are compatible with those found in the previous analysis~\cite{Lenz:2023rlq}.

Compared with the case in which the QCD penguin contributions are neglected, including these operators leads to an enhancement of the resulting bounds. Given the significant uncertainties, this reduces the gap with the experimental measurement. The effect is particularly pronounced when NLO Wilson coefficients are used. As discussed above, however, the use of NLO Wilson coefficients should be regarded as illustrative, since the accuracy of our analysis does not extend to a fully consistent NLO treatment.
It is therefore important to stress that the results obtained here should be interpreted as maximal values rather than predictions. A quantitative prediction within LCSR requires control over the relative strong phases and the associated $U$-spin-breaking corrections, which remains challenging and is beyond the accuracy of our analysis. Gaining such control is therefore an essential direction for future work. Nevertheless, our analysis indicates that the inclusion of the previously neglected QCD penguin contributions to the LCSR calculation can have a significant impact on the SM expectation for direct CP violation in charm, highlighting the need for more comprehensive studies of these effects.
%%%%%%%%%%%%%%%%%%%%%%%%%%%%%%%%%%%%%%%%%%%%%%%%%
%%%%%%%%%%%%%%%%%%%%%%%%%%%%%%%%%%%%%%%%%%%%%%%%%
%%%%%%%%%%%%%%%%%%%%%%%%%%%%%%%%%%%%%%%%%%%%%%%%%
\section{Discussion and conclusions}
\label{sec:conclusion}
In this work we have performed a systematic analysis of the contributions of the current-current and QCD penguin operators to the decay amplitudes of the singly Cabibbo suppressed decays $D^0 \to \pi^+ \pi^-$ and $D^0 \to K^+ K^-$, within the framework of LCSR using pion and kaon LCDAs and at leading order in $\alpha_s$. 
Our study extends and complements the previous analysis~\cite{Lenz:2023rlq}, which considered only the current-current operators, and provides the first LCSR determination of the matrix elements of the QCD penguin operators for the above modes. 
Although QCD penguin operators are usually neglected in phenomenological analyses because of their small Wilson coefficients, their contribution is proportional to the CKM-suppressed combination $\lambda_b$ and therefore enters the part of the decay amplitude relevant for direct CP violation.
Moreover, the QCD penguin operators mix under renormalisation with the current-current operators. While the penguin contractions of the current-current operators have been estimated using LCSR in~\cite{Khodjamirian:2017zdu}, a non-perturbative determination of the matrix elements of the QCD penguin operators themselves has been lacking. Our analysis provides a first step towards assessing their size and quantifying their potential impact on charm CP violation.

We have adapted to the charm sector the LCSR framework originally developed for non-leptonic $B \to \pi \pi$ decays~\cite{Khodjamirian:2000mi}, introducing an auxiliary momentum in the correlation function to avoid contributions from parasitic cuts in the dispersion relations. This setup allows us to consistently include contributions beyond the tree-level topology, in particular annihilation contributions. For the current-current operators, these contributions are numerically suppressed because they arise only through subleading twist-four LCDAs. As a result, our matrix elements are in good agreement with the previous LCSR analysis~\cite{Lenz:2023rlq}, with central values approximately $10\%$ smaller. This agreement provides a non-trivial cross-check of the LCSR framework. The corresponding branching-ratio predictions for $D^0\to K^+K^-$ and $D^0\to\pi^+\pi^-$ are also consistent with previous determinations~\cite{Lenz:2023rlq} and the HFLAV averages~\cite{HFLAV:2024ctg}.  While our results exhibit a smaller degree of $U$-spin breaking, mainly because $m_s$ corrections are not included in our analysis, they nevertheless reproduce the sizeable difference experimentally observed between the two decay modes. 

The main new result of this work is the determination of the matrix elements of the QCD penguin operators at leading order in $\alpha_s$. We find that the matrix elements of $Q_5$ and $Q_6$ can be substantially larger than those of the current-current operators. This enhancement is a consequence of their $(V-A)\otimes(V+A)$ Dirac structure, which allows them to couple directly to the chirally enhanced pseudoscalar light-meson LCDAs. 

In particular, we identify two sources of enhancement. On the one hand, the quark-condensate contributions give a sizeable effect. Although these terms formally correspond to twist-five and twist-six contributions, when combined with twist-two and twist-three LCDAs, see~\cite{Khodjamirian:2002pk} for details, they arise already at tree level, whereas the remaining terms in the sum rule originate from one-loop diagrams. They are therefore enhanced by a relative factor of $16\pi^2$. On the other hand, an annihilation-type contribution originating from the $\bar u u$ component of these penguin operators receives contributions already from twist-three LCDAs, yielding an effect of comparable magnitude to the quark-condensate term.

By contrast, for the remaining operators, quark-condensate contributions are absent at the leading order considered, while the annihilation contributions are numerically suppressed, arising only from subleading twist-four LCDAs. 
It would be interesting to investigate whether the inclusion of three-particle LCDAs can modify this pattern. In particular, it would be important to determine whether the corresponding twist-four contributions cancel the residual effects found here or whether these contributions persist once the complete set of higher-twist corrections is included.
 
Despite the enhancement of the QCD penguin matrix elements, their impact on the branching ratios remains negligible because of the suppression by both the Wilson coefficients and the CKM factor $\lambda_b$. Their impact on direct CP violation, however, is considerably more pronounced. Including the QCD penguin operators enhances the ratios $|{\cal P}_{\pi\pi}/{\cal A}_{\pi\pi}|$ and $|{\cal P}_{KK}/{\cal A}_{KK}|$ relative to the case in which only penguin contractions of the current-current operators are retained, by factors of approximately two and four, respectively, depending on whether LO or NLO Wilson coefficients are used. We stress that the latter case should be regarded as illustrative since our analysis does not reach full NLO accuracy.

Assuming maximal relative strong phases, the inclusion of the QCD penguins can enhance the SM value of $|\Delta a_{\rm CP}^{\rm dir}|$, and within the quoted
uncertainties, reduce the gap with the experimental measurement. The resulting maximal value depends on the Wilson coefficients used, with the NLO coefficients leading to a higher bound. 
At the same time, the present uncertainties remain sizeable, in particular those associated with missing higher-order and power corrections and with the assumption of QHD. While a robust quantitative prediction within LCSR will ultimately require improved control over these theoretical uncertainties and reliable information on the relative strong phases, our results indicate that previously neglected QCD penguin contributions can have a significant impact on the SM expectation for charm CP violation and represent a further step towards a comprehensive description of hadronic $D^0$ decays within this framework. 
%%%%%%%%%%%%%%%%%%%%%%%%%%%%%%%%%%%%%%%%%%%%%%%%%
%%%%%%%%%%%%%%%%%%%%%%%%%%%%%%%%%%%%%%%%%%%%%%%%%
%%%%%%%%%%%%%%%%%%%%%%%%%%%%%%%%%%%%%%%%%%%%%%%%%
\section*{Acknowledgments}
We thank Alexander Khodjamirian and Aleksey Rusov for helpful discussions, as well as for carefully reading the manuscript and providing insightful comments. We are also grateful to Alexander Lenz for his valuable comments on the manuscript.
MLP wishes to thank Luiz Vale Silva for interesting discussions that have partly inspired this work. 
AM gratefully acknowledges the hospitality of the CERN Theory Department, where part of this work was carried out.
The research of
MLP is funded by the European Union’s Horizon Europe
Research and Innovation Programme under the Marie
Sk{\l}odowska-Curie grant agreement No.\ 101204923. 
The work of AM is supported by the Deutsche Forschungsgemeinschaft
(DFG, German Research Foundation) under Germany's Excellence Strategy,
EXC 3107 -- Project ID 533766364 and under grant 396021762-TRR 257.
\appendix 
\section{Supplementary material}
\label{sec:app1}
In this appendix, we collect additional definitions and expressions used throughout the paper. The matrix elements of the pseudoscalar $D$-meson current,
$j_5^D(x)=i m_c \bar u\gamma_5 c$, and the axial-vector current
$j_\mu^L(x)=\bar u\gamma_\mu\gamma_5 q$, with $q={d,s}$ for
$L={\pi,K}$, between the vacuum and the corresponding meson states are given by
\begin{equation}
    \langle 0 | j_5^D (x)| D(p) \rangle = m_D^2 f_D e^{- i p \cdot x}\,, \qquad
    \langle 0 | j_\mu^L (x)| L(p) \rangle = i f_L p_\mu e^{- i p \cdot x}\,,
\end{equation}
where $f_D$ and $f_L$ denote the corresponding meson decay constants.  

For the expansion of the vacuum matrix elements of quark bilinears in terms of the local quark condensate, we use, see e.g.~\cite{Khodjamirian:2020btr},
\begin{align}
\langle  0 | \bar q_\alpha^i(x) q_\beta^j(0)|0 \rangle = \frac{1}{4 N_c} \langle \bar q q\rangle \delta_{i j} \delta_{\beta \alpha}  + i \frac{m_q}{16 N_c} \langle \bar q q\rangle \delta_{ij}  x^\mu (\gamma_\mu)_{\beta\alpha}\,,\\
\langle 0 | \bar q_\alpha^i(0) q_\beta^j(x)|0\rangle = \frac{1}{4 N_c} \langle \bar q q\rangle \delta_{i j} \delta_{\beta \alpha}  - i \frac{m_q}{16 N_c} \langle \bar q q\rangle \delta_{ij} x^\mu (\gamma_\mu)_{\beta\alpha}\,.
\end{align}
Here, $q = u,d,s$ and $\langle \bar q q \rangle \equiv \langle 0 |\bar q q | 0 \rangle$ denotes the quark-condensate density. Using the Gell-Mann-Oakes-Renner relation~\cite{Gell-Mann:1968hlm}, we express the quark condensates in terms of the chirally enhanced parameters $\mu_\pi$ and $\mu_K$ as
\begin{align}
\langle \bar q q \rangle &= - (1/2) f_\pi^2 \mu_\pi\,, \qquad q = u,d \,,
\label{eq:QC-qq-relation}
\\[2mm]
\langle \bar q q \rangle &= - (1/2) f_K^2 \mu_K\,, \qquad q = s \,.
\label{eq:QC-ss-relation}
\end{align}
The first relation holds up to corrections of
${\cal O}(m_u^2,m_d^2)$, while the second receives corrections of
${\cal O}(m_u^2,m_s^2)$, which are parametrically larger. Hence, the relation between the strange-quark condensate and $\mu_K$ is expected to receive larger $SU(3)_f$-breaking corrections than the corresponding relation for the up and down-quark condensate.

To derive the imaginary
part of the functions entering the OPE results, we use
\begin{equation}
{\rm Im}  \frac{1}{(x - i \varepsilon)^n} = \pi \frac{(-1)^{n-1}}{(n-1)!} \delta^{(n-1)}(x) \,, \quad n \geq 1\,,
\end{equation}
where $\delta^{(n-1)}(x)$ indicates the $(n-1)$--derivative of the delta function with respect to its argument. Moreover, the analytic continuation of the logarithm function at
negative values of the argument is defined as
\begin{equation}
 \ln (-x) = \ln |x| - i \pi \theta(x)\,,
\end{equation}
where $\theta(x)$ is the Heaviside function.
%%%%%%%%%%%%%%%%%%%%%%%%%%%%%%%%%%%
%%%%%%%%%%%%%%%%%%%%%%%%%%%%%%%%%%%
%%%%%%%%%%%%%%%%%%%%%%%%%%%%%%%%%%%
\section{Relations between matrix elements}
\label{sec:app1p}
In this appendix, we derive the relations between the matrix elements of the colour-singlet and colour-rearranged operators in eqs.~\eqref{eq:Q1} - \eqref{eq:Q5}. For simplicity, in the discussion of the LC-OPE in section~\ref{sec:OPE} we have focused on operators with the structure of eq.~\eqref{eq:operator}, namely, for the $\pi^+ \pi^-$ final state, on the operators $O = \{ Q_2^d,Q_3,Q_5\}$. The corresponding results for the other matrix elements follow by using the identity in eq.~\eqref{eq:Fierz_id}. For instance, in the case of the colour-singlet operator $Q_1^d$, applying a Fierz transformation together with the identity in eq.~\eqref{eq:Fierz_id}, we obtain
\begin{equation}
C_1 Q_1^d + C_2 Q_2^d = \left( C_2 + \frac{C_1}{N_c}\right) Q_2^d + 2 C_1 \tilde Q^d_2\,,
\end{equation}
where the colour-octet operator is defined as in eq.~\eqref{eq:octet_op}. The matrix element of $Q_2^d$ receives contributions from the topologies $\rm P_I$ and $\rm P_{II}$ in eqs.~\eqref{eq:T1}, \eqref{eq:T2}. The corresponding contributions to the matrix element of the colour-octet operator at leading order in $\alpha_s$ read, respectively, up to overall prefactors,
\begin{equation}
\bar d^i(0) \Gamma_2^{\tilde Q_2^d}t^a_{ij} S_{0}^{(d)}(-y) \delta_{jm} \gamma_\mu \gamma_5 S_0^{(u)}(y) \delta_{ml} \Gamma_1^{\tilde Q_2^d} t^a_{lk} S_0^{(c)} (-x) \delta_{kn} \gamma_5 u^n(x)\,, \quad {\rm (P_I)}
\end{equation}
and 
\begin{equation}
\bar d^i(0) \Gamma_2^{\tilde Q_2^d} t^a_{ij} S_0^{(d)}(-y) \delta_{jm} \gamma_\mu \gamma_5 u^m(y) \, {\rm Tr} \bigg[\Gamma_1^{\tilde Q_2^d} t^a_{lk} S_0^{(c)}(-x) \delta_{kn} \gamma_5 S_0^{(u)}(x) \delta_{nl}\bigg]\,. \quad {\rm (P_{II})}
\end{equation}
The traceless of the $SU(3)$ generators, ${\rm Tr}(t^a) = 0$ implies that the contribution from the topology $\rm P_{II}$ to the matrix element of the colour-octet operator vanishes while the property $t^a_{ij} t^a_{jm} = C_F \delta_{im}$, with $C_F = (N_c^2-1)/(2 N_c)$, implies that the contribution of $\rm P_{I}$ is proportional to the one in eq.~\eqref{eq:T1} up to a colour factor, i.e.
\begin{align}
C_1 \langle Q_1^d\rangle_{\pi \pi} + C_2 \langle Q_2^d \rangle_{\pi \pi} & = \left( C_2 + \frac{C_1}{N_c} \right) \bigg( \langle Q_2^d \rangle_{\pi \pi}^{\rm P_I} + \langle Q_2^d \rangle_{\pi \pi}^{\rm P_{II}}\bigg) + C_1 \left(\frac{N_c^2-1}{N_c} \right)\langle Q_2^d \rangle_{\pi \pi}^{\rm P_{I}}
\\[2mm]
& = \left(C_1 + \frac{C_2}{N_c} \right) N_c \langle Q_2^d \rangle_{\pi \pi}^{\rm P_{I}} + \left( C_2 + \frac{C_1}{N_c} \right)  \langle Q_2^d \rangle_{\pi \pi}^{\rm P_{II}}\,.
\end{align}
Matching the coefficients of $C_1$ and $C_2$ then gives, at the order considered here,
\begin{equation}
    \langle Q_1^d \rangle_{\pi \pi}^{\rm P_{I}} = N_c \langle Q_2^d \rangle_{\pi \pi}^{\rm P_{I}}\,, \qquad \langle Q_1^d \rangle_{\pi \pi}^{\rm P_{II}} = \frac{1}{N_c} \langle Q_2^d \rangle_{\pi \pi}^{\rm P_{II}}\,.
\end{equation}
Similarly, one can derive the relations between the colour-singlet and colour-rearranged QCD penguin operators. In particular, for $k = 4,6$: 
\begin{equation}
\begin{cases}
\langle Q_k \rangle_{\pi \pi}^{\rm X} = N_c \langle Q_{k-1} \rangle_{\pi \pi}^{\rm X}\,, & \rm X = P_I , P_{III}\,,
\\[2mm]
\langle Q_k \rangle_{\pi \pi}^{\rm X} = \frac{1}{N_c} \langle Q_{k-1} \rangle_{\pi \pi}^{\rm X}\,, & \rm X = P_{II} , P_{IV}\,,
\end{cases}
\end{equation}
while 
\begin{equation}
    \langle Q_6 \rangle_{\pi \pi}^{\rm QC} = N_c  \langle Q_5 \rangle_{\pi \pi}^{\rm QC}\,.
\end{equation}
%%%%%%%%%%%%%%%%%%%%%%%%%%%%%%%%%%%
%%%%%%%%%%%%%%%%%%%%%%%%%%%%%%%%%%%
%%%%%%%%%%%%%%%%%%%%%%%%%%%%%%%%%%%
\section{Expressions of the OPE coefficients}
\label{sec:app2}
The non-vanishing coefficient functions
$c_{\phi,n,\mathrm I}^{Q_2^d}$
defined in eq.~(\ref{eq:FI_OPE}) are given by
\begingroup
\allowdisplaybreaks[4]
\begin{align}
c_{\phi_{2\pi},1,\mathrm I}^{Q_2^d}
&=
\frac{m_c }{24\pi^2u}
\left(
P^2-(q-k)^2-m_\pi^2
\right),
\\[2mm]
c_{\phi_{3\pi}^{p},1,\mathrm I}^{Q_2^d}
&=
\frac{\mu_\pi}{24\pi^2u}
\left[
u\left(P^2-m_\pi^2\right)
+\bar u\,(q-k)^2
\right],
\\[2mm]
c_{\phi_{3\pi}^{\sigma},1,\mathrm I}^{Q_2^d}
&=
\frac{\mu_\pi}{144\pi^2u^2}
\left[
2u\left(P^2-m_\pi^2\right)
+(1-2u)(q-k)^2
\right],
\\[2mm]
\notag c_{\phi_{3\pi}^{\sigma},2,\mathrm I}^{Q_2^d}
&=
\frac{\mu_\pi}{144\pi^2u^3}
\Bigg[
u\left(m_c^2-m_\pi^2u^2\right)
\left(P^2-m_\pi^2\right)
\\[2mm]
&-
\left(
m_\pi^2u^2\bar u
+m_c^2(1+u)
\right)
(q-k)^2
\Bigg],
\\[2mm]
c_{\phi_{4\pi},3,\mathrm I}^{Q_2^d}
&=
-\frac{m_c^2}{2u^2}\,
c_{\phi_{2\pi},1,\mathrm I}^{Q_2^d},
\\[2mm]
c_{\psi_{4\pi},2,\mathrm I}^{Q_2^d}
&=
-\frac{m_c}{u\mu_\pi}\,
c_{\phi_{3\pi}^{p},1,\mathrm I}^{Q_2^d}.
\end{align}
The independent non-vanishing coefficient functions
$c_{\phi,n,{\rm X}}^{O}$ in
eq.~(\ref{eq:F_II_IV_OPE}) are given by
\begin{align}
c_{\phi_{2\pi},1,\mathrm{II}}^{Q_3}
&=
-\frac{m_c N_c^2}{48\pi^2u}
\frac{(m_c^2-p^2)^2}{(p^2)^2}
\left(p^2+m_\pi^2\right),
\\[2mm]
c_{\phi_{4\pi},2,\mathrm{II}}^{Q_3}
&=
-\frac{1}{4u}\,
c_{\phi_{2\pi},1,\mathrm{II}}^{Q_3},
\\[2mm]
c_{\psi_{4\pi},2,\mathrm{II}}^{Q_3}
&=
-\frac{m_c N_c^2}{24\pi^2u^2}
\frac{(m_c^2-p^2)^2}{(p^2)^2}
\left(
P^2+\bar u\,p^2-u\,m_\pi^2
\right).
\\[2mm]
%\end{align}
%\begin{align}
c_{\phi_{2\pi},1,\mathrm{III}}^{Q_3}
&=
\frac{m_c N_c}{48\pi^2\bar u}
\frac{(m_c^2-p^2)^2}{(p^2)^2}
\left(p^2+m_\pi^2\right),
\\[2mm]
c_{\phi_{4\pi},2,\mathrm{III}}^{Q_3}
&=
-\frac{1}{4\bar u}\,
c_{\phi_{2\pi},1,\mathrm{III}}^{Q_3},
\\[2mm]
c_{\psi_{4\pi},2,\mathrm{III}}^{Q_3}
&=
-\frac{m_c N_c}{24\pi^2\bar u^2}
\frac{(m_c^2-p^2)^2}{(p^2)^2}
\left(
P^2+u\,p^2-\bar u\,m_\pi^2
\right).
\\[2mm]
c_{\phi_{3\pi}^{p},1,\mathrm{III}}^{Q_5}
&=
-\frac{N_c\mu_\pi}{12\pi^2\bar u}
\frac{(m_c^2-p^2)^2}{p^2},
\\[2mm]
c_{\phi_{3\pi}^{\sigma},1,\mathrm{III}}^{Q_5}
&=
-\frac{1}{6\bar u}\,
c_{\phi_{3\pi}^{p},1,\mathrm{III}}^{Q_5},
\\[2mm]
c_{\phi_{3\pi}^{\sigma},2,\mathrm{III}}^{Q_5}
&=
\frac{P^2-m_\pi^2\bar u^2}{\bar u}\,
c_{\phi_{3\pi}^{\sigma},1,\mathrm{III}}^{Q_5}.
\end{align}
The coefficients for topology~IV are related to those of topology~II by
\begin{align}
c_{\phi_{2\pi},1,\mathrm{IV}}^{Q_i}
&=
-\left.
c_{\phi_{2\pi},1,\mathrm{II}}^{Q_i}
\right|_{u\to\bar u},
\\[2mm]
c_{\phi_{4\pi},2,\mathrm{IV}}^{Q_i}
&=
-\left.
c_{\phi_{4\pi},2,\mathrm{II}}^{Q_i}
\right|_{u\to\bar u},
\\[2mm]
c_{\psi_{4\pi},2,\mathrm{IV}}^{Q_i}
&=
\left.
c_{\psi_{4\pi},2,\mathrm{II}}^{Q_i}
\right|_{u\to\bar u},
\end{align}
where $i=3,5$. The remaining coefficient functions satisfy
\begin{align}
c_{\phi,n,\mathrm{II}}^{Q_2^d}
&=
c_{\phi,n,\mathrm{II}}^{Q_3}
=
c_{\phi,n,\mathrm{II}}^{Q_5},
\\[2mm]
c_{\phi,n,\mathrm{IV}}^{Q_5}
&=
c_{\phi,n,\mathrm{IV}}^{Q_3}.
\end{align}
The non-vanishing coefficient functions
$c_{\phi,n,\mathrm{QC}}^{Q_5}$
entering
eq.~\eqref{eq:FqQCI-OPE} are
\begin{align}
c_{\phi_{2\pi},1,\mathrm{QC}}^{Q_5}
&=
\frac{m_c^2+m_\pi^2u^2}{3u^2},
\\[2mm]
c_{\phi_{3\pi}^{p},1,\mathrm{QC}}^{Q_5}
&=
\frac{2m_c\mu_\pi}{3u},
\\[2mm]
c_{\phi_{4\pi},1,\mathrm{QC}}^{Q_5}
&=
\frac{1}{12u^2},
\\[2mm]
c_{\phi_{4\pi},2,\mathrm{QC}}^{Q_5}
&=
\frac{m_c^2-m_\pi^2u^2}{u}\,
c_{\phi_{4\pi},1,\mathrm{QC}}^{Q_5},
\\[2mm]
c_{\phi_{4\pi},3,\mathrm{QC}}^{Q_5}
&=
-\frac{m_c^2}{2u^2}\,
c_{\phi_{2\pi},1,\mathrm{QC}}^{Q_5},
\\[2mm]
c_{\psi_{4\pi},1,\mathrm{QC}}^{Q_5}
&=
-\frac{2}{3u},
\\[2mm]
c_{\psi_{4\pi},2,\mathrm{QC}}^{Q_5}
&=
\frac{m_c^2}{u}\,
c_{\psi_{4\pi},1,\mathrm{QC}}^{Q_5}.
\end{align}
\endgroup
Corresponding results for $D^0 \to K^+ K^-$ are obtained with the replacements $m_\pi \to m_K$ and $\mu_\pi \to \mu_K$, together with the proper replacement of the pion LCDAs with the kaon LCDAs.
\newpage

\bibliographystyle{JHEP}
\bibliography{References}

\end{document}